\documentclass[aps,prb,reprint,amsmath,amssymb,floatfix,nofootinbib,showkeys]{revtex4-2}

\usepackage{graphicx}
\usepackage{mathtools}
\usepackage{microtype}
\usepackage{placeins}
\usepackage{xcolor}
\usepackage{booktabs}
\usepackage{array}
\usepackage{bm}
\usepackage{xspace}
\usepackage[caption=false]{subfig}
\usepackage[
  colorlinks=true,
  linkcolor=blue,
  citecolor=magenta,
  urlcolor=cyan
]{hyperref}
\usepackage[normalem]{ulem}

\graphicspath{{_comparisons/}}

\newcommand{\mnfe}{MnFe$_2$O$_4$\xspace}
\newcommand{\mnzn}{(Mn$_{0.5}$Zn$_{0.5}$)Fe$_2$O$_4$\xspace}
\newcommand{\fe}{Fe$_3$O$_4$\xspace}
\newcommand{\Wcyc}{W_{\mathrm{cyc}}}

\begin{document}

\title{Modeling Dynamic Magnetic Response of Spinel Soft Ferrites with the Steepest-Entropy-Ascent Quantum Thermodynamics Formalism}

\author{Deepak Dhariwal}
\affiliation{Department of Materials Science and Engineering, Virginia Tech, Blacksburg, VA 24060, USA}
\author{William T. Reynolds, Jr.}
\affiliation{Department of Materials Science and Engineering, Virginia Tech, Blacksburg, VA 24060, USA}
\author{Michael R. von Spakovsky}
\affiliation{Department of Mechanical Engineering, Virginia Tech, Blacksburg, VA 24060, USA}

\date{\today}

\begin{abstract}
Selecting soft ferrites for alternating-field applications requires balancing magnetic response, dissipation, nonlinearity, and heating. We use a field-driven steepest-entropy-ascent quantum thermodynamic (SEAQT) model to compare the electron, phonon, and magnon responses of Fe$_3$O$_4$, MnFe$2$O$4$, and (Mn${0.5}$Zn${0.5}$)Fe$_2$O$_4$ within a common first-principles framework. The model treats spatially uniform longitudinal relaxation and neglects domain-wall motion, transverse rotation, resonance, eddy-current effects, and heat removal. We use the electron and phonon relaxation parameters $\tau_e=0.05$ ps and $\tau_p=3$ ps for all three materials, with effective longitudinal magnon relaxation parameters of 500, 200, and 85 ps, respectively. These literature-motivated values are model inputs, not fits to measured losses. Within this formalism, the Mn--Zn ferrite gives the largest peak longitudinal magnetization change, retains its response best at high frequency, and shows the largest work per cycle and peak-to-peak magnon temperature change, whereas MnFe$_2$O$_4$ gives the largest normalized imaginary susceptibility, $\chi''$, peak. Thus, no single ferrite ranks highest across all metrics. Instead, the preferred material depends on the property, frequency, field amplitude, and cation configuration of interest. The model, therefore, provides a spectrum- and dynamics-resolved screening tool for engineering comparison of ferrites rather than a prediction of total core loss.
\end{abstract}


\maketitle

\section{Introduction}
\label{sec:intro}

Iron-oxide-containing soft magnets called ferrite cores combine ferrimagnetic response with comparatively high electrical resistivity, a combination that has long made them useful in inductors, transformers, filters, and other high-frequency magnetic components~\cite{Goldman2006,Stoppels1996,Thakur2020,Snoek1948}. Their usable operating range, however, is not controlled by one intrinsic material property. Permeability, dynamic loss, saturation, thermal stability, and frequency response depend jointly on chemistry, cation distribution, microstructure, specimen geometry, and excitation conditions~\cite{Dobak2022,Harris2012}.

This interdependent character is especially well documented in Mn--Zn ferrites. Complex permeability contains contributions from domain-wall motion and magnetization rotation, and the relative importance of these contributions changes with frequency, bias, and microstructure~\cite{Visser1984,Lebourgeois1996,Tsutaoka1999,Dobak2022,Wu2024}. Loss-separation studies likewise show that hysteretic, rotational or residual, spin-damping, and eddy-current contributions do not remain in fixed proportion as the operating conditions change~\cite{Beatrice2006,Sun2011,Bertotti1988,Dobak2022}. These intricacies are reflected in conventional engineering models. Steinmetz-type laws and their waveform-aware extensions remain valuable precisely because they collapse the behavior of a manufactured core into compact design relations~\cite{Steinmetz1892,Reinert2001,Muehlethaler2012}. Modern power-electronics models have also introduced explicit frequency-dependent relaxation terms for ferrites when quasistatic hysteresis alone is insufficient, emphasizing that relaxation is already an engineering concern even when its microscopic origin is not resolved spectrally~\cite{Luo2019}. However, the parameters inferred from such models
do not, by themselves, reveal how the measured magnetic performance of a material arises from the underlying electronic, vibrational, and magnetic excitation spectra.

A spectrum-based comparison is especially attractive for spinel ferrites because chemical substitution alters several aspects of the problem simultaneously. Within a fixed crystal structure, replacing and redistributing Fe, Mn, and Zn alters near-Fermi electronic states, exchange pathways, spin-wave bandwidths, and the distribution of lattice vibrations. The accompanying cation arrangement is not a secondary bookkeeping choice. Measurements on Mn--Zn ferrites show that A- and B-sublattice occupancies can vary with preparation history, and that mixed or partially inverted distributions are physically accessible~\cite{Sakurai2008}. An engineering model that aims to connect the chemistry to the dynamic response should, therefore, be able to distinguish a compositional change from a cation-configurational change.

The spectral inputs needed for such a comparison were established previously using one internally consistent first-principles workflow for \fe (magnetite), \mnfe, and \mnzn~\cite{Dhariwal2026JCTC}. That density functional theory (DFT) work generated electronic, phonon, and magnon densities of states (DOS) while preserving a common computational reference across compositions and representative cation arrangements. The calculated spectra differ in ways that are directly relevant to relaxation. Magnetite retains pronounced near-Fermi electronic weight in one spin channel, whereas the Mn-containing systems are more strongly depleted near the Fermi level in the studied configurations. The magnon spectra are even more composition sensitive: \fe has the largest calculated spin-wave bandwidth of the set, redistributing Mn between A and B sites softens \mnfe, and Zn-containing configurations compress the magnon spectrum further and place more weight at lower energies. The phonon bandwidth changes less dramatically, although substitution and cation placement redistribute substantial vibrational weight within that bandwidth~\cite{Dhariwal2026JCTC}. These differences provide a controlled microscopic basis for exploring how ferrites with the same spinel structure but different underlying electronic spectra relax under an alternating magnetic field.

A field-driven SEAQT model was subsequently developed for \fe to establish a fundamental description of the dynamic relaxation of electron, phonon, and magnon spectra in an alternating magnetic field~\cite{Dhariwal2026SEAQT}. In that formulation, a longitudinal magnetic field shifts the magnon eigenenergy ladder and drives a redistribution of magnon occupations. Electrons, phonons, and magnons are coupled by the constraints on the evolution of the thermodynamic state rather than by a fitted magnetic-loss relationship. The formulation separates magnetic work from entropy production, yields a coupled complex susceptibility in the small-signal regime, and extends naturally to finite-amplitude nonlinear response. Its principal restriction is deliberate: it describes longitudinal quasiparticle-population relaxation within one homogeneous magnetic region. It does not contain domain walls, transverse precession, macroscopic reversal, specimen-scale electromagnetic diffusion, or a heat-interaction boundary.

This article uses the same model~\cite{Dhariwal2026SEAQT} as a common dynamical framework to compare relaxation of the ferrite spectra reported in Ref.~\cite{Dhariwal2026JCTC}. The purpose is not to repeat the spectral study and not to reproduce the SEAQT derivation. Instead, the objective is to compare how material-specific spectra and prescribed kinetic scales jointly affect observables that are relevant to engineering applications: frequency-dependent magnetization, complex susceptibility, relaxational work, entropy production, higher-harmonic distortion, and internal subsystem temperatures. A second objective is to separate chemistry-level differences from changes caused only by cation distribution within the same nominal composition.

The aforementioned observables do not support a single overall ranking of the materials. A material with a larger change in magnetization can absorb more energy in absolute terms even if its normalized imaginary susceptibility, $\chi''$, peak is relatively small. Similarly, a shorter magnetic relaxation parameter does not necessarily mean that the material reaches its maximum energy loss earlier, because the electron, phonon, and magnon populations exchange energy and respond together. A material that shows little harmonic distortion at one frequency may also behave more nonlinearly at another. The purpose of this comparison is, therefore, not to establish a universal performance metric, but to provide a more detailed view of one intrinsic relaxation mechanism. This information can then be combined with models of domain behavior, electromagnetic effects, and heat interaction to evaluate material performance at the component scale.

The model presented and the comparison made in the following sections are limited to that of a defect-free, single-crystal systems in the presence of a time varying magnetic field. The focus is solely on the behavior of the electron, phonon, and magnon populations relative to each other in the absence of any domain walls. The systems are also assumed to be thermally isolated from their environment with no induced electrical current and, thus, no eddy losses. All of these features could be added via the SEAQT formalism but are left for future work. The remainder of the paper is organized as follows. Section \ref{sec:model} provides a brief description of the material energy eigenstructures described in detail in \cite{Dhariwal2026JCTC} and of the principal equations of the SEAQT model derived in \cite{Dhariwal2026SEAQT}. Section \ref{sec:protocol} then outlines what comparisons are made in the following section of the materials' non-equilibrium thermodynamic behavior, while Section \ref{sec:results} discusses the results of this behavior. This is followed in Section \ref{sec:limitations} by a discussion of the any limitations and validations. The paper then concludes in Section \ref{sec:conclusions} with some final remarks.

\section{Materials and model}
\label{sec:model}

\subsection{Ferrite systems and spectral inputs}
\label{subsec:materials}

The comparison presented here uses the excitation spectra reported in Ref.~\cite{Dhariwal2026JCTC}. Magnetite is the Fe-rich reference. Two \mnfe cation arrangements are included: configuration 1 places Mn on the tetrahedral A sublattice, while configuration 2 distributes Mn between A and B sites. Three \mnzn arrangements are used: configuration A places Mn and Zn on A sites; configuration B keeps Zn on A sites and places Mn on B sites; and configuration C keeps Zn on A sites while distributing Mn between A and B sites. These labels are used only where the comparison files identify the configuration explicitly.

The configuration-specific electronic, magnon, and phonon spectra used here are those reported and discussed in Ref.~\cite{Dhariwal2026JCTC}. The present work uses those spectra as inputs to the nonequilibrium relaxation calculation rather than reproducing their detailed spectral comparison.

The electronic, phonon, and magnon populations are all included in the thermodynamic state. The magnetic field acts directly on the magnon energies, but the common energy constraint redistributes some of this energy to the other two populations. Consequently, a change in the electronic or phonon DOS can affect the magnetic response even though the external field is introduced through the magnon sector. This is why the comparison is described throughout as spectrum- and dynamics-resolved rather than as a magnetic-relaxation-time comparison alone.

\subsection{Field-driven SEAQT model}
\label{subsec:minimal}

The DOS-based and hypoequilibrium foundations of the SEAQT description are developed in Refs.~\cite{Beretta2014,Li2016Hypo,Li2016Relations,Kim2017,Li2018Transport,Li2018Mesoscopic,Yamada2019Methods,Yamada2019Spin,BerettaRay2026Hypo}. In particular, Beretta, Ray, and von Spakovsky provide a recent operator-level treatment of hypoequilibrium-state evolution and its invariant-manifold structure~\cite{BerettaRay2026Hypo}. The details behind the magnetic model for ferrites --- including construction of the thermodynamic system, definitions of the state variables, constraints, and magnetic work as well as derivations of the equation of motion and the small-signal range --- are provided in reference~\cite{Dhariwal2026SEAQT}.  For simplicity, only the equations needed to interpret the model inputs and the plotted observables are presented in the following expressions.

The imposed longitudinal magnetic field strength is
\begin{equation}
    H(t)=H_b+H_0\cos\omega t,
    \label{eq:field}
\end{equation}
where $H_b$ is the strength of a bias field, $H_0$ a sinusoidal amplitude, and $\omega\, (=2\pi f)$ the angular frequency. The field shifts a magnon level of zero-field energy $\varepsilon_j^m$ to
\begin{equation}
\epsilon_j^m(t)=\varepsilon_j^m+\mu_0\,\mu_m\,H(t),
\label{eq:magnonshift}
\end{equation}
where $\mu_0$ is the vacuum permeability and $\mu_m>0$ is the effective ordered magnetic moment removed by adding one magnon. Electron and phonon levels are not shifted by the magnetic field in this reduced model.
For electrons and phonons, which are not shifted directly by the magnetic field in this reduced model, $\epsilon_j^e=\varepsilon_j^e$ and $\epsilon_j^p=\varepsilon_j^p$.

The species-level relaxation laws~\cite{Dhariwal2026SEAQT} are
\begin{align}
    \dot y_j^e&=-\frac{y_j^e-\beta \varepsilon_j^e-\nu}{\tau_e},\nonumber\\
    \dot y_j^p&=-\frac{y_j^p-\beta \varepsilon_j^p}{\tau_p},\nonumber\\
    \dot y_j^m&=-\frac{y_j^m-\beta \epsilon_j^m(t)}{\tau_m}.
    \label{eq:compacteom}
\end{align}
Here $y_j^k$ is the dimensionless population coordinate for level $j$ of species $k\in\{e,p,m\}$. The $y_j^k$ determine the mean population of species $k$ through the Fermi map (when $k$ corresponds to electrons) or Bose statistics (when $k$ corresponds to phonons or magnons). The common multiplier $\beta(t)$ is conjugate to the total instantaneous excitation-energy constraint. It is determined so that the irreversible redistribution of the electron, phonon, and magnon populations leaves the total excitation energy unchanged at a given instant of time, with the field-shifted magnon energies $\epsilon_j^m(t)$ used in evaluating that total. The multiplier $\nu(t)$ separately enforces conservation of total electron number. The quantities $\tau_e$, $\tau_p$, and $\tau_m$ are dynamic relaxation parameters introduced to scale the relaxation of respective populations to observed time frames. Magnon number is not imposed as a conserved property so that the longitudinal magnetization can relax.

For the species-affine states used to report subsystem temperatures, each population has a nonequilibrium inverse-temperature coordinate $\beta_k$. The corresponding plotted temperature is
\begin{equation}
    T_k=\frac{1}{k_b\beta_k}, \qquad k\in\{e,p,m\},
    \label{eq:temperatures}
\end{equation}
where $k_b$ is Boltzmann's constant. Under a changing magnetic field, the magnon state also carries a field-generated affinity so $T_m$ is a useful coordinate but not, by itself, a complete description of the driven magnon population.

The magnetic observable, the magnetization, $M$, follows the magnon population density, $\langle n\rangle_m$, such that
\begin{equation}
    M=M_{\mathrm{sat}}-\mu_m \langle n\rangle_m,
    \label{eq:magnetization}
\end{equation}
where $M_{\mathrm{sat}}$ is the ordered reference magnetization. Increasing the magnon population, therefore, reduces the longitudinal magnetization on the chosen ordered branch. Here the ordered branch denotes the fixed ferrimagnetic orientation used to construct the spin-wave spectrum. The modeled magnetization can change longitudinally about that reference state through changes in the magnon population, but switching to the oppositely oriented ferrimagnetic state is not included.

For one period of Eq.~(\ref{eq:field}), the relaxational work density is
\begin{equation}
    \Wcyc=\mu_0\oint H\,dM.
    \label{eq:work}
\end{equation}
This is the energy transferred through the modeled longitudinal electron--phonon--magnon relaxation contribution. The SEAQT equation of motion also provides a separate entropy-production rate,
\begin{equation}
    \frac{\dot \sigma}{k_b}=\sum_{k,j} r_j^k\left(\Delta_j^k\right)^2\ge 0,
    \label{eq:entropy}
\end{equation}
Here $\Delta_j^k$ is the departure of level $j$ from its instantaneous constrained target and $r_j^k>0$ is its kinetic/occupation response weight~\cite{Dhariwal2026SEAQT}. Explicitly, $\Delta_j^k=y_j^k-\beta \epsilon_j^k(t)-\nu\delta_{ke}$ and $r_j^k=g_j^kA_j^k/\tau_k$, with $g_j^k$ the spectral degeneracy, $A_j^k$ the occupation-fluctuation factor, $\epsilon_j^k(t)$ the instantaneous level energy, and $\delta_{ke}$ equal to one only for electrons~\cite{Dhariwal2026SEAQT}. The cycle-integrated entropy production in subsequent plots is $\sigma_{\mathrm{cyc}}=\int_{\mathrm{cyc}}\dot\sigma\,dt$. Eqs.~\eqref{eq:work} and \eqref{eq:entropy} describe related but different aspects of the system evolution: energy absorption and irreversibility, respectively.

In the small-signal limit, the model gives the coupled susceptibility expressed as~\cite{Dhariwal2026SEAQT}
\begin{equation}
    \chi(\omega)=\mu_0\,\mu_m^2\,\beta_0\left[
    \frac{B_{NN}^m}{1+\mathrm{i}\,\omega\,\tau_m}
    -\frac{(B_{EN}^m)^2}{(1+\mathrm{i}\,\omega\,\tau_m)^2G(\omega)}
    \right],
    \label{eq:chi}
\end{equation}
with
\begin{equation}
    G(\omega)=\frac{C_e}{1+\mathrm{i}\,\omega\,\tau_e}
    +\frac{B_{EE}^p}{1+\mathrm{i}\,\omega\,\tau_p}
    +\frac{B_{EE}^m}{1+\mathrm{i}\,\omega\,\tau_m}.
    \label{eq:G}
\end{equation}
Here $\mathrm{i}^2=-1$ denotes the imaginary unit. The spectral-level index is $j$. Here, $\beta_0=(k_bT_0)^{-1}$ is the equilibrium inverse temperature at reference temperature $T_0$. $B_{NN}^m$ is the equilibrium magnon occupation-fluctuation sum, $B_{EN}^m$ is its first energy-weighted moment of the field-shifted magnon levels, and $B_{EE}^m$ is its second energy-weighted moment of the field-shifted magnon levels. $B_{EE}^p$ is the corresponding phonon second energy moment, and $C_e$ is the electronic fixed-number energy variance. Their numerical values are determined by the material-specific spectra and equilibrium occupations~\cite{Dhariwal2026SEAQT}. The first term of Eq.~\eqref{eq:chi} is a direct magnon response, whereas the second contains the electron--phonon--magnon feedback required by energy conservation. A single Debye pole is recovered only when that feedback is negligible.

Expressing the susceptibility as a complex number $\chi=\chi'-\mathrm{i}\chi''$, with in-phase (real) component $\chi'$ and out-of-phase (imaginary) component $\chi''$, the linear-response work per cycle is
\begin{equation}
\Wcyc^{(2)}=\pi\,\mu_0\,H_0^2\,\chi''  \;\;.
\label{eq:linearwork}
\end{equation}
At finite amplitude, the magnetization is no longer restricted to the fundamental frequency. The quantity $|M_3|/|M_1|$ used below is the magnitude of the third Fourier harmonic divided by that of the fundamental. It is employed as a measure of distortion, whereas Eq.~\eqref{eq:work} remains the energy metric. This distinction becomes important when comparing the results across materials of cyclic magnetic fields with large amplitudes.

\subsection{Kinetic inputs and their physical basis}
\label{subsec:kinetic_provenance}

The SEAQT model assigns one effective relaxation parameter to each excitation population: $\tau_e$ for electrons, $\tau_p$ for phonons, and $\tau_m$ for magnons. In the species-uniform dynamic metric (or scheme), the parameters $\tau_k,\; k \in \{e, p, m\}$ set the time scales on which each of the species eigenlevel populations relax toward the constrained SEAQT target. Because that target is determined by the coupled conservation constraints, these times characterize the rates of coupled population relaxation rather than lifetimes of isolated microscopic scattering events. The electron relaxation parameter is fixed at $\tau_e=0.05$ ps and that for phonons at $\tau_p=3$ ps for each of the ferrite compositions. The magnon relaxation parameter, $\tau_m$, is material dependent: 500 ps for \fe, 200 ps for \mnfe, and 85 ps for \mnzn. The values are collected in Table~\ref{tab:kinetic_inputs} for reference before the numerical results are discussed.

\begin{table}[t]
\caption{Species-level relaxation parameters used in all calculations.}
\label{tab:kinetic_inputs}
\centering
\begingroup
\setlength{\tabcolsep}{9pt}
\begin{tabular}{lccc}
\toprule
Material & $\tau_e$ (ps) & $\tau_p$ (ps) & $\tau_m$ (ps) \\
\midrule
\fe   & 0.05 & 3 & 500 \\
\mnfe & 0.05 & 3 & 200 \\
\mnzn & 0.05 & 3 & 85 \\
\bottomrule
\end{tabular}
\endgroup
\end{table}

These parameters should be interpreted as coarse-grained dynamic parameters of the SEAQT state-space metric, not as unique lifetimes of one microscopic scattering event. This distinction is particularly important for the electron and phonon values. The electronic coordinate in the model describes redistribution of the electronic population over the supplied electronic eigenstructure. It is not an electron-hopping time extracted from dc transport. Likewise, $\tau_p$ is an effective phonon-population relaxation parameter and not the lifetime of one Raman mode or a single electron--phonon scattering event.

The choice of $\tau_e=0.05$ ps places the electronic sector in the ultrafast regime expected for electronic redistribution in magnetite. Time-resolved resonant X-ray diffraction has reported loss of electronic order in Fe$_3$O$_4$ on a time scale of about $30\pm30$ fs, with the initially excited electrons cascading into lower-energy electronic excitations within still shorter times~\cite{Pontius2018}. Ultrafast optical measurements on epitaxial Fe$_3$O$_4$ likewise show sub-100-fs demagnetization and an electronic reflectivity response that develops within a few tenths of a picosecond~\cite{Lu2022Ultrafast}. These experiments do not measure the SEAQT $\tau_e$ directly, but they place a 50-fs species-level electronic relaxation scale in a physically reasonable range for fast electronic redistribution.

The choice of $\tau_p=3$ ps represents a slower lattice-sector scale. Ultrafast electron diffraction on magnetite resolves a first structural response over approximately 0.7--3.2 ps and a second stage that develops after about 3.2 ps, reflecting distinct charge--lattice and subsequent lattice-reorganization processes~\cite{Wang2022UED}. Optical pump--probe work also observes electron--phonon-associated relaxation on sub-picosecond-to-picosecond time scales in Fe$_3$O$_4$~\cite{Lu2022Ultrafast}. A 3-ps phonon metric, therefore, sits in the experimentally observed picosecond window for lattice redistribution. It should not be read as a universal phonon lifetime.

The same $\tau_e$ and $\tau_p$ are deliberately retained for all three ferrites. Equivalent composition-resolved ultrafast measurements are not available with sufficient consistency to assign separate values without introducing a second layer of poorly constrained material dependence. Holding these two times fixed makes the comparison controlled: differences arising from the electron and phonon sectors enter through their material-specific density of states (DOS), whereas the imposed composition dependence in the dynamic metric is restricted to the magnetic population time. This assumption is useful for comparison but is not a claim that electron and phonon relaxation are truly identical in all ferrites. A future calibration study should relax that constraint once comparable measurements are available.

Unlike $\tau_e$ and $\tau_p$, which are held fixed here to maintain a controlled comparison, $\tau_m$ is allowed to vary among the ferrites. The effective magnetic-population dynamics need not be universal because changes in cation chemistry and site occupancy modify exchange interactions, magnetic anisotropy, spin--orbit and spin--lattice coupling, and the microscopic pathways by which a nonequilibrium magnon population relaxes.  These processes are not determined by the magnon DOS alone. Their unresolved dynamic effect is represented separately by $\tau_m$ in the SEAQT model. 

Available experiments do not provide a unique observable that can be identified one-to-one with this longitudinal SEAQT parameter. They, nevertheless, constrain its physically reasonable scale. Hsia \textit{et al.} measured Fe$_3$O$_4$ magnetization recovery increasing from about 250 to 350 ps with nanocrystal size and found substantially stronger spin–lattice relaxation at surfaces than interiors~\cite{Hsia2009}. A recent study by Jeyaram and Joseyphus directly considers Fe$_3$O$_4$, MnFe$_2$O$_4$, and Mn–Zn ferrites and reports calculated spin–lattice $T_1$ values spanning $2.80$–$5.18\times10^{-10}$ s and spin–spin $T_2$ values spanning $2.18$–$4.02\times10^{-11}$ s~\cite{Jeyaram2025}. Resonance and linewidth measurements provide additional evidence that magnetic relaxation is strongly material- and environment-dependent \cite{Lee2022FMR,Wang2022MZFO}. These quantities are not SEAQT $\tau_m$ values, but they should behave similarly and they demonstrate that $\tau_m$ values are composition dependent and fall within the tens-to-hundreds-of-picoseconds range.
 
On this basis, $\tau_m=500$, 200, and 85 ps are specified before the present comparison as nominal slow, intermediate, and fast longitudinal magnetic-population relaxation times for Fe$_3$O$_4$, MnFe$_2$O$_4$, and (Mn$_{0.5}$Zn$_{0.5}$)Fe$_2$O$_4$, respectively. The 500-ps value places magnetite near the slow end of the experimentally observed few-hundred-picosecond magnetic-relaxation range, whereas 85 ps places the mixed Mn--Zn ferrite in the faster tens-to-hundreds-of-picoseconds regime.  The 200-ps MnFe$_2$O$_4$ value is an intermediate model input and is the least directly constrained of the three.  None of these values is fitted to the susceptibility, work, entropy-production, or harmonic results reported below.  They should, therefore, be regarded as literature-bounded nominal dynamic inputs whose uncertainty must ultimately be assessed by sensitivity analysis and experimental calibration.

The resulting hierarchy, $\tau_e\ll\tau_p\ll\tau_m$, is therefore a deliberate coarse graining of fast electronic redistribution, intermediate lattice relaxation, and slower longitudinal magnetic-population relaxation. The three magnetic values are not fitted to the calculated work, susceptibility, or harmonic curves. The 200-ps value for \mnfe remains the least directly constrained composition-specific value and should be read as an intermediate model-level time within the experimentally supported sub-nanosecond range. These limitations are carried into the interpretation rather than hidden in the parameter table.

\subsection{Scope of the comparison}
\label{subsec:scope}

The magnetic system is a homogeneous, single domain, collinear magnetic region. Domain-walls, transverse magnetization components, anisotropy-axis rotation, specimen-scale electric currents, and Maxwell-field redistribution are excluded from the state space. The calculated $M$--$H$ trajectories should, thus, be interpreted as longitudinal relaxation paths about one ordered branch rather than as macroscopic hysteresis loops. Correspondingly, $\Wcyc$ is one intrinsic quasiparticle-relaxation contribution to loss. It is not the total loss of a manufactured magnetic core.

This distinction is important because measured ferrite permeability and loss are known to contain domain-wall, rotational, residual, spin-damping, and eddy-current contributions whose relative importance changes with microstructure, frequency, field amplitude, and bias~\cite{Visser1984,Lebourgeois1996,Beatrice2006,Tsutaoka1999,Dobak2022,Wu2024}. The present calculation is intended to isolate one piece of that larger problem so that its chemistry dependence can be studied without convoluting all the loss mechanisms into empirical coefficients.

The thermodynamic model also lacks a heat sink. For this reason, positive magnetic work done on the system raises the internal excitation energy, and the three populations exchange that energy with one another.  The model does not exchange thermal energy with a substrate, winding, package, coolant, or ambient environment. The temperature curves are, thus, internal thermodynamic diagnostics rather than device steady-state temperatures.

\section{Numerical comparison}
\label{sec:protocol}

\subsection{Source data}
\label{subsec:compared_data}

The numerical results are generated from the DFT-generated pseudo-eigenstructures presented in Ref.~\cite{Dhariwal2026JCTC} and the field-driven equations summarized in Sec.~\ref{sec:model} but dervide in Ref. \cite{Dhariwal2026SEAQT}. The sets of compared data contain family-level results that provide one representative trace for each of the \fe, \mnfe, and \mnzn materials. These do not identify which Mn-containing cation configuration was used, so no A/B-site assignment is made to the curves. In some cases, results for different configurations are compared, and the configuration-specific conclusions are drawn only from plots that label configurations 1--2 or A--C. 

\begin{figure*}[t]
    \centering
    \subfloat[]{%
    \includegraphics[width=0.47\textwidth]{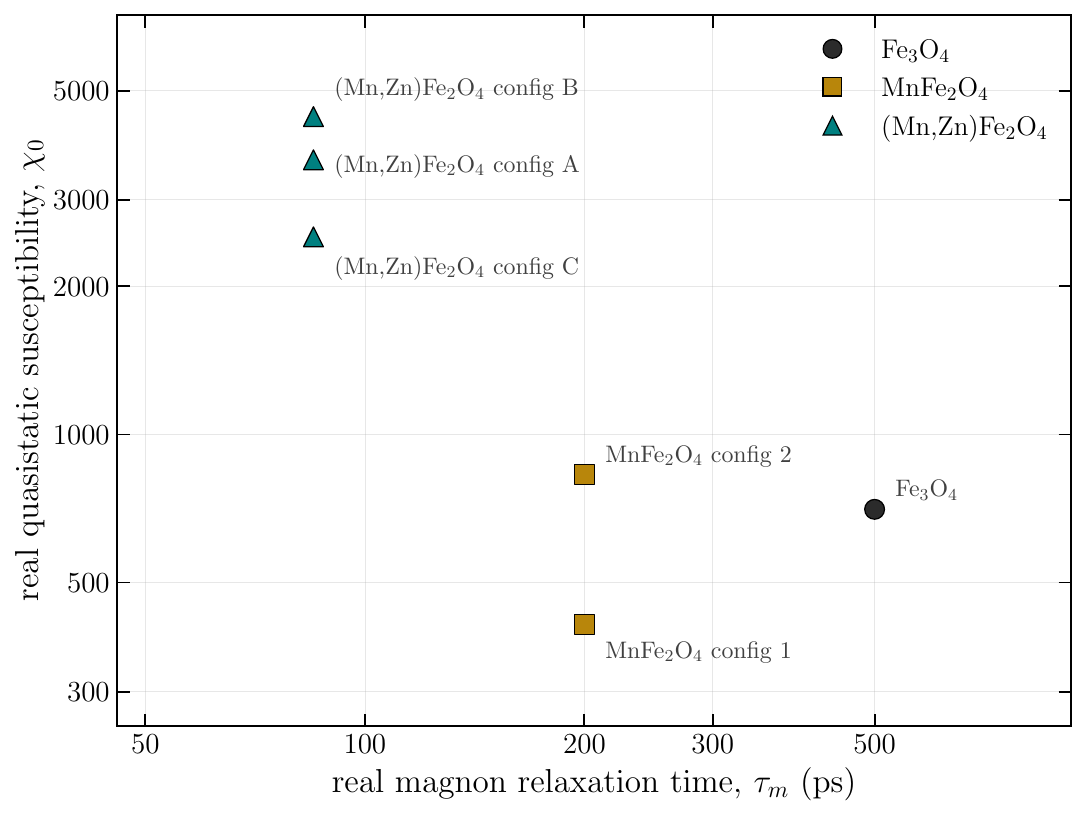}}
    \hfill
    \subfloat[]{%
    \includegraphics[width=0.47\textwidth]{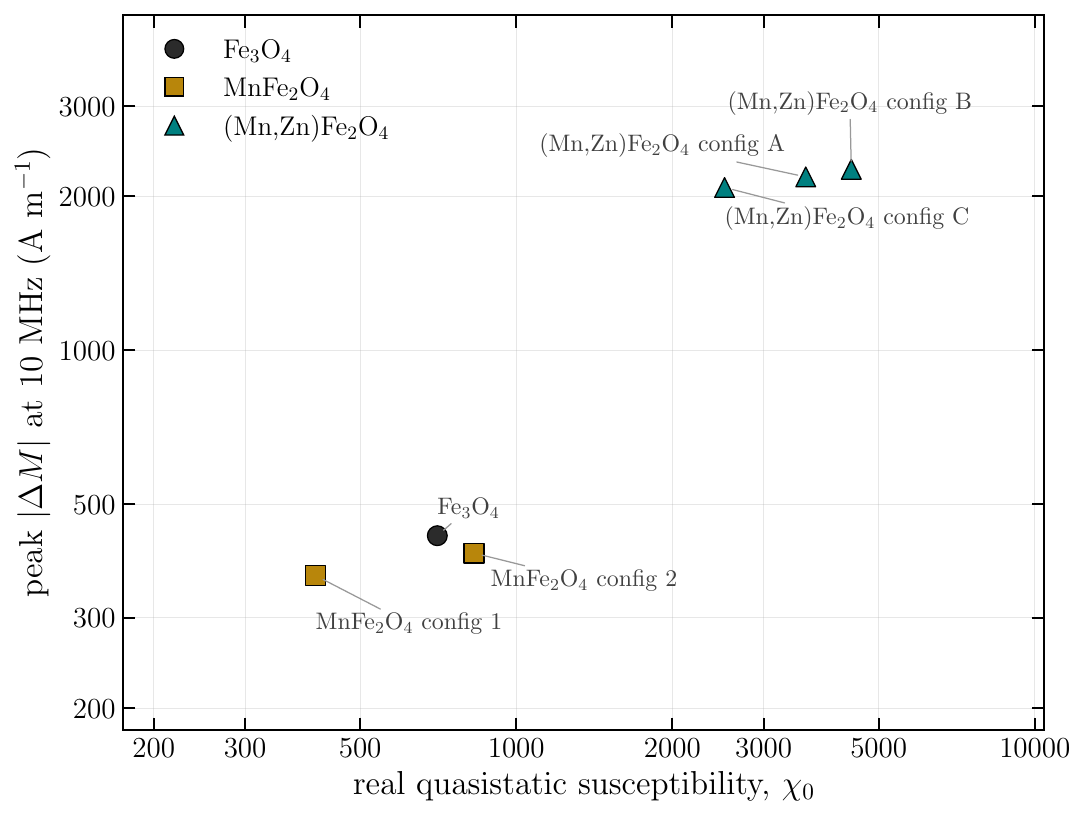}}
    \caption{Maps indicating the range of susceptibilities and magnetization changes: (a) prescribed magnetic relaxation parameter, $\tau_m$, versus zero-frequency in-phase susceptibility $\chi_0$ and (b) $\chi_0$ versus the peak longitudinal magnetization change $|\Delta M|_{\mathrm{peak}}$ at 10 MHz.}
    \label{fig:material_maps}
\end{figure*}

 Two maps are presented in Fig.~\ref{fig:material_maps}(a) and (b) to identify the range of several important input parameters. In these figures, $\chi_0\equiv\lim_{\,\omega\rightarrow0}\,\chi'(\omega)$ denotes the  zero-frequency limit of the in-phase susceptibility and $\Delta M=M-M_{\mathrm{ref}}$ is the plotted longitudinal magnetization change relative to reference value $M_{\mathrm{ref}}$. The peak longitudinal magnetization change is defined as $|\Delta M|_{\mathrm{peak}}=\max_{t\in\mathrm{cycle}}|M(t)-M_{\mathrm{ref}}|$. Here, $M_{\mathrm{ref}}$ is the magnetization of the equilibrium reference state at the stated initial temperature and bias field, established before the sinusoidal field protocol is applied. Thus, $M_{\mathrm{ref}}$ is the baseline from which the driven longitudinal change is measured and is distinct from the ordered reference magnetization $M_{\mathrm{sat}}$ in Eq.~\eqref{eq:magnetization}. The peak quantity measures the largest longitudinal magnetization change reached during a field cycle and gives an absolute measure of the modeled response. It is not by itself a measure of dissipation, which also depends on the phase relation between $M$ and $H$. The quantity $\chi_0\equiv\lim_{\omega\to0}\chi'(\omega)$ establishes the low-frequency response scale against which finite-frequency suppression can be compared.

Fig.~\ref{fig:material_maps}(a) places the prescribed magnetic relaxation parameter beside the zero-frequency limit of the in-phase susceptibility, $\chi_0=\lim_{\omega\rightarrow0}\chi'(\omega)$. The map makes clear that these are independent inputs rather than two representations of one parameter. The Mn--Zn points combine the shortest $\tau_m$ with the largest $\chi_0$, whereas the two MnFe$_2$O$_4$ configurations occupy an intermediate dynamic range but noticeably different susceptibility values. Fig.~\ref{fig:material_maps}(b) then shows that the 10-MHz peak $|\Delta M|$ follows the susceptibility scale much more closely than it follows $\tau_m$ alone. This is the first indication that the magnitude of the low-frequency response and the frequency range over which that response is retained should be considered separately.

\subsection{Operating range and outputs}
\label{subsec:conditions}

The family-level frequency sweeps extend from the MHz range to 5 GHz, depending on the observable. Configuration-resolved $M$--$H$ comparisons emphasize 10 and 100 MHz together with 1 and 5 GHz frequencies. Finite-amplitude work, entropy, and harmonic results use the dimensionless drive amplitude index $\xi$, with larger $\xi$ indicating a larger amplitude of the imposed driving magnetic field.  In the available comparison data, $\xi$ is a dimensionless, monotonic drive-amplitude index: increasing $\xi$ denotes increasing imposed sinusoidal-field amplitude. No material-independent conversion from $\xi$ to a physical $H_0$ is assumed in the comparisons below.

The principal outputs are peak $|\Delta M|$, the real and imaginary susceptibility components $\chi'$ and $\chi''$, cycle work $\Wcyc$, cycle-integrated entropy production $\sigma_{\mathrm{cyc}}$,  the harmonic ratio $|M_3|/|M_1|$, and the electron, phonon, and magnon temperatures. 
The family-level temperature comparison reports the electron, phonon, and magnon peak-to-peak within-cycle temperature changes, together with the mean magnon temperature over successive cycles~\cite{Stoppels1996,Ott2003,Liu2008,Tokatlidis2018,Tsutaoka1999,Harris2012,Kawano2000,Sun2011}.



\section{Results and discussion}
\label{sec:results}

\subsection{Frequency-dependent response}
\label{subsec:dynamic}

\begin{figure}[t]
    \centering
    \includegraphics[width=\columnwidth]{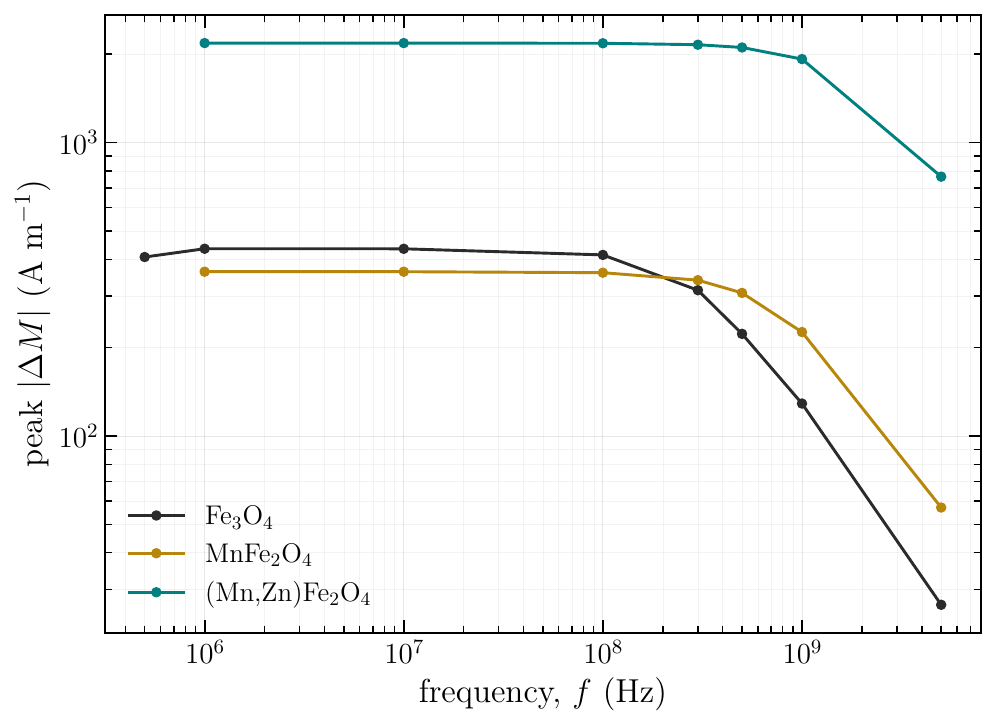}
    \caption{Peak longitudinal magnetization change $|\Delta M|_{\mathrm{peak}}$ versus cyclic drive frequency $f$ (Hz).}
    \label{fig:family_mrange}
\end{figure}

Fig.~\ref{fig:family_mrange} compares the  peak longitudinal magnetization change across the three ferrite families. At the low-frequency end, the peak $\Delta M$ for \fe and for \mnfe are similar, whereas that of \mnzn is almost an order of magnitude larger.  As frequency increases, the magnitude of the longitudinal magnetization change $|\Delta M|_{\mathrm{peak}}$ decreases sharply. This drop in magnetic response with increasing frequency occurs first for \fe ferrite at a frequency of $10^8$ Hz, then \mnfe, and finally \mnzn at $10^9$ Hz. \mnfe retains more of its magnetic response at high frequencies than \fe, but \mnzn maintains the largest response over the entire frequency range, and it retains a substantial peak $\Delta M$ even at the highest frequency investigated.

The ordering in Fig.~\ref{fig:family_mrange} reflects two distinct effects: $\chi_0$ sets how large the low-frequency magnetization response is for a given small field amplitude, whereas the relaxation parameters and DOS-dependent coupling determine how rapidly that response is suppressed as the drive frequency increases. The shorter $\tau_m$ assigned to the Mn-containing materials gives their magnetic population more opportunity to follow a faster field, but the much larger magnitude of the Mn--Zn response cannot be attributed to $\tau_m$ alone. Fig.~\ref{fig:material_maps}(b) already shows that the larger zero-frequency susceptibility $\chi_0$ of the Mn--Zn ferrite produces a larger low-frequency magnetization change for the same small applied-field amplitude. Eq.~\eqref{eq:chi} further shows that the frequency at which $\chi'$ begins to decrease and the rate at which it decreases thereafter depend on the electron, phonon, and magnon contributions to the common energy constraint rather than on $\tau_m$ alone. The family ordering, therefore, reflects both the magnitude of the low-frequency magnetic response and the finite rates of the coupled population redistribution.

The upper end of the frequency sweep is used as a diagnostic extension of the model response rather than as a claim about the usual operating range of Mn--Zn power ferrites. Mn--Zn power ferrites are commonly designed for kHz-to-low-MHz operation, while specialized high-frequency ferrites and permeability studies extend into the MHz regime~\cite{Stoppels1996,Ott2003,Liu2008,Tokatlidis2018,Tsutaoka1999,Harris2012}. Measurements in the MHz range also show that the relative contributions to measured ferrite loss change with frequency and temperature~\cite{Kawano2000,Sun2011}. The present calculations are extended into the GHz range so that the high-frequency suppression of the modeled longitudinal electron--phonon--magnon relaxation response and material-dependent crossovers can be observed within one continuous comparison. The GHz results should, thus, be interpreted as a diagnostic extension of this intrinsic relaxation calculation, not as a statement that conventional Mn--Zn power cores are normally operated at several GHz.

\begin{figure*}[t]
\centering
\subfloat[]{%
\includegraphics[width=0.48\textwidth]{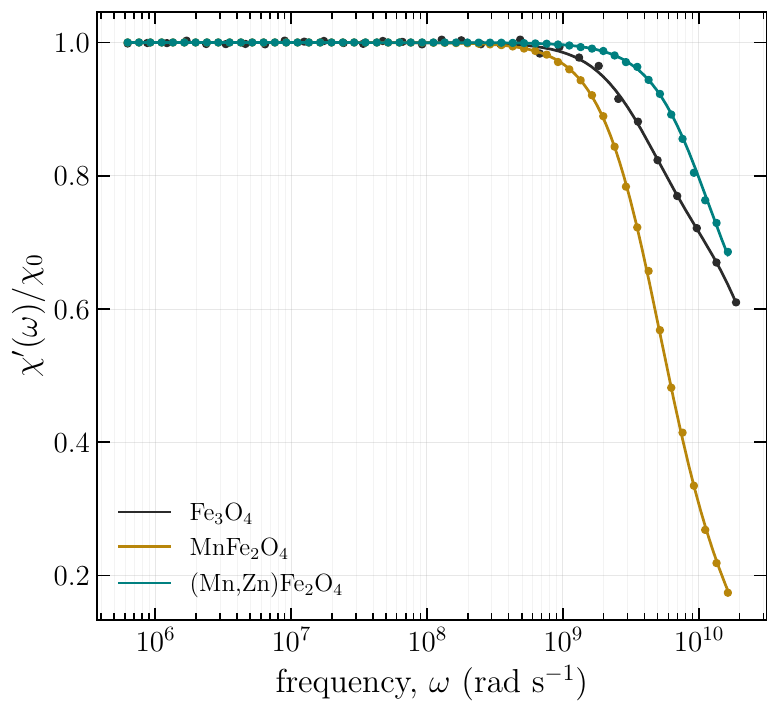}}
\hfill
\subfloat[]{%
\includegraphics[width=0.48\textwidth]{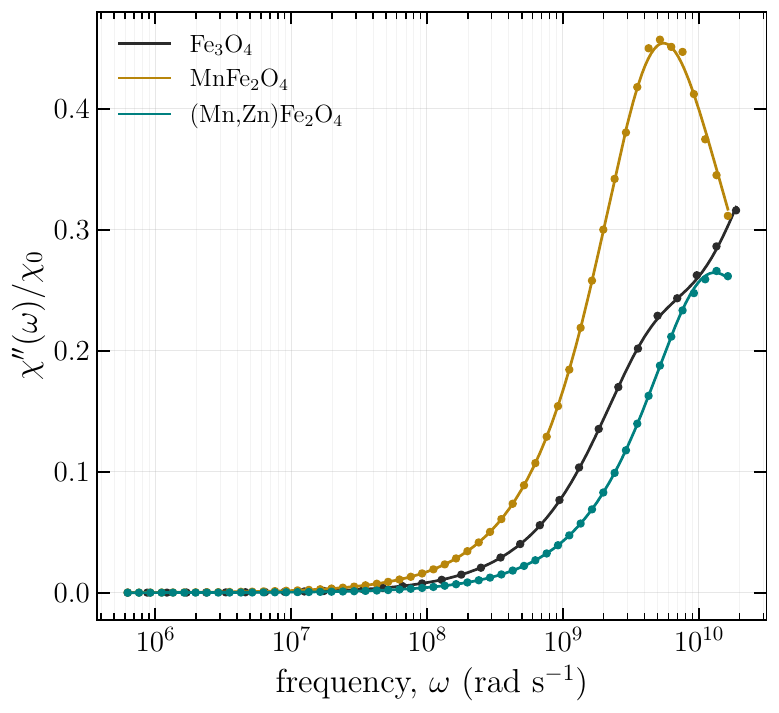}}
\caption{Normalized small-signal susceptibility for the three ferrite families: (a) $\chi'(\omega)/\chi_0$ and (b) $\chi''(\omega)/\chi_0$.}
\label{fig:family_chi}
\end{figure*}

To compare how much of each material's low-frequency in-phase magnetic response remains as the drive is accelerated, Fig.~\ref{fig:family_chi}(a) plots $\chi'(\omega)/\chi_0$. A value near unity means that the in-phase susceptibility remains close to its zero-frequency value. A decrease below unity quantifies suppression of that in-phase response as the drive frequency increases. The out-of-phase component $\chi''$ in panel (b) is considered separately because it quantifies the quadrature response associated with phase lag and magnetic work. The three $\chi'/\chi_0$ curves remain near unity at low angular frequency and then decrease as the field is driven faster than the coupled populations can follow. The onset and breadth of that decrease differ materially. \mnzn  preserves a larger fraction of its zero-frequency in-phase susceptibility $\chi_0$ farther into the high-frequency range, whereas \fe loses normalized response earlier. The representative \mnfe curve turns over even earlier.

Fig.~\ref{fig:family_chi}(b) shows the part of the small-signal magnetization response that is out of phase with the applied field. A single-relaxation-time Debye model also has an out-of-phase component, but it predicts a specific one-time-scale frequency dependence set by one characteristic relaxation time. The SEAQT result differs because the directly driven magnon population exchanges energy with the electron and phonon populations through the common energy constraint. Consequently, the rise, breadth, and largest value of $\chi''(\omega)/\chi_0$ are not determined by $\tau_m$ alone.

For the engineering comparison, three features of Fig.~\ref{fig:family_chi}(b) are more informative than the peak position by itself: the angular-frequency range in which the out-of-phase response becomes appreciable, the interval over which it remains appreciable, and its largest normalized magnitude. The Fe$_3$O$_4$ and MnFe$_2$O$_4$ curves begin to rise from their near-zero baseline at lower angular frequency than the Mn--Zn curve, whereas the Mn--Zn response is shifted toward higher angular frequency. MnFe$_2$O$_4$ attains the largest plotted value of $\chi''/\chi_0$, Fe$_3$O$_4$ reaches a somewhat smaller value, and the Mn--Zn ferrite has the smallest maximum of the three representative curves. The different frequency dependences cannot be inferred from $1/\tau_m$ alone because Eq.~\eqref{eq:chi} also contains $\tau_e$, $\tau_p$, and energy-weighted contributions from the electron, phonon, and magnon densities of states through $G(\omega)$.

\begin{figure}[t]
\centering
\includegraphics[width=\columnwidth]{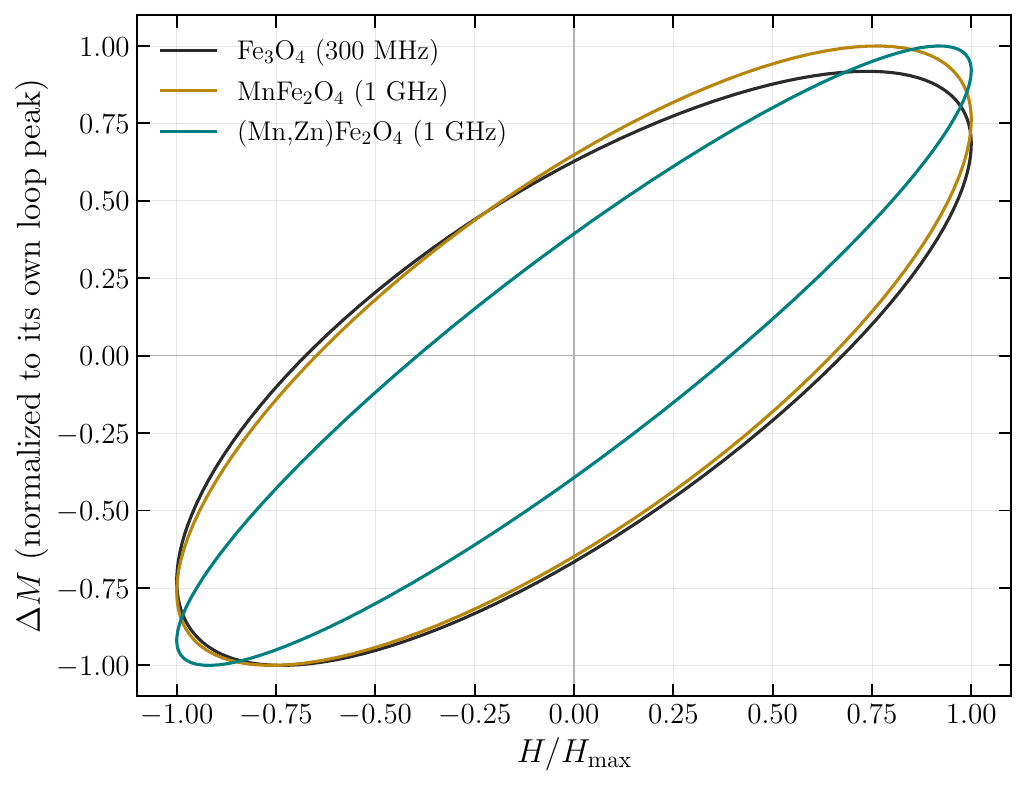}
\caption{Normalized $M$--$H$ loop shapes near the dispersive regime of each family.}
\label{fig:near_corner}
\end{figure}

  Fig.~\ref{fig:near_corner} is used to compare the phase-lag geometry of the three responses after the large material-to-material differences in field and magnetization amplitude have been divided out. For an approximately linear sinusoidal response, the relative in-phase and out-of-phase susceptibility components determine the orientation and opening of the $M$--$H$ ellipse. Thus, the normalized loop shape provides a time-domain visual counterpart to the susceptibility comparison in Fig.~\ref{fig:family_chi}. The Mn--Zn loop is narrower after normalization than the Fe$_3$O$_4$ and MnFe$_2$O$_4$ loops, consistent with its smaller normalized out-of-phase response over the representative dispersive condition shown. The three curves are evaluated at different frequencies---300 MHz for Fe$_3$O$_4$ and 1 GHz for MnFe$_2$O$_4$ and the Mn--Zn ferrite---so this figure compares representative loop geometry rather than same-frequency performance. Because both axes are normalized, the area enclosed by Fig.~\ref{fig:near_corner} is dimensionless and must not be interpreted as the absolute cycle work. The absolute energy density per cycle is obtained from the unnormalized trajectory via Eq.~\eqref{eq:work}.

 Taken together, Figs.~\ref{fig:material_maps}--\ref{fig:near_corner} establish three distinct observations. First, the low-frequency response magnitudes differ strongly among the ferrite families, with the Mn--Zn ferrite having the largest $\chi_0$ and the largest low-frequency $|\Delta M|_{\mathrm{peak}}$. Second, the normalized in-phase and out-of-phase susceptibilities change over different angular-frequency ranges, demonstrating that the frequency dependence cannot be inferred from $\tau_m$ alone. Third, after the field and magnetization amplitudes are normalized, the loop shapes are more similar than the absolute magnetization changes. Thus normalized phase-lag geometry and absolute response magnitude should not be treated as interchangeable measures. The next section, therefore, compares the absolute energy density transferred during a cycle.

\subsection{Work and entropy production}
\label{subsec:workentropy}

\begin{figure*}[t]
\centering
\subfloat[]{%
\includegraphics[width=0.48\textwidth]{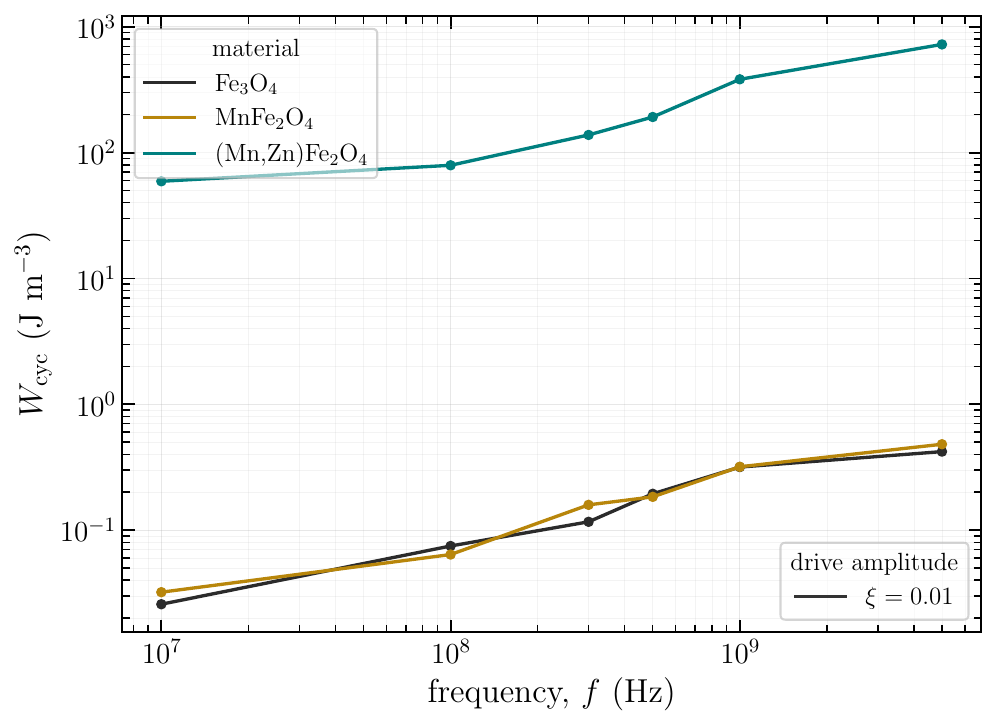}}
\hfill
\subfloat[]{%
\includegraphics[width=0.48\textwidth]{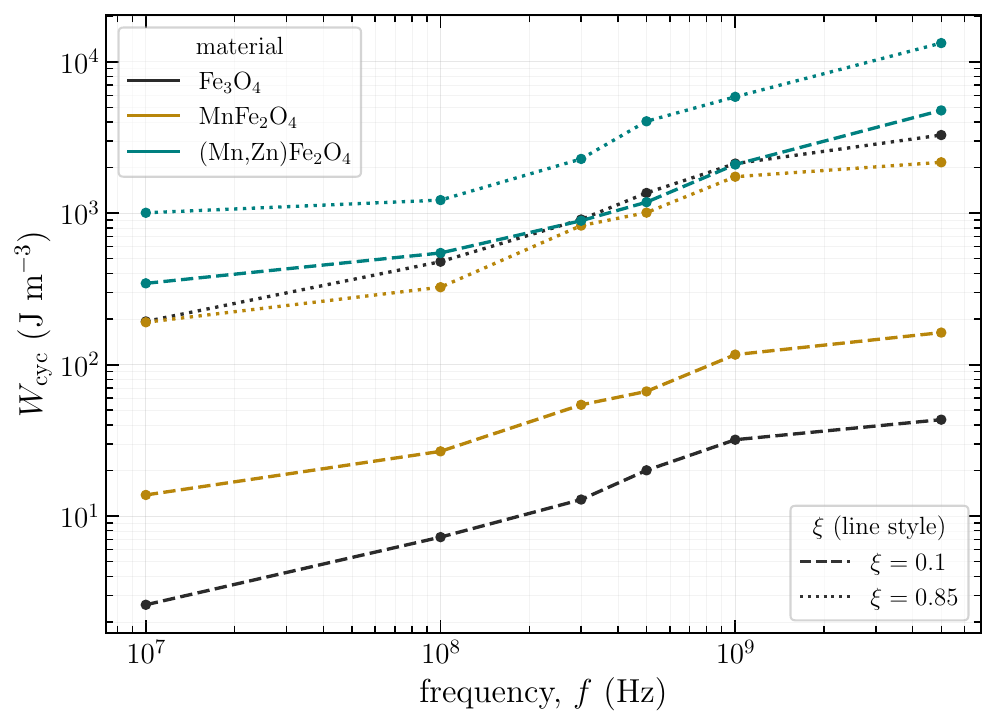}}
\caption{Cycle-work density, $W_{\mathrm{cyc}}$, versus cyclic drive frequency for Fe$_3$O$_4$, MnFe$_2$O$_4$, and (Mn$_{0.5}$Zn$_{0.5}$)Fe$_2$O$_4$. (a) Weak-drive comparison at $\xi=0.01$. (b) Corresponding comparison at $\xi=0.1$ and $0.85$. The material is identified by color; in panel (b), the drive-amplitude index is identified by line style.}
\label{fig:family_work}
\end{figure*}

Fig.~\ref{fig:family_work} separates the weak-drive comparison from the two larger drive amplitudes so that the material ordering is visible without overlap between amplitude groups. At $\xi=0.01$ [panel (a)], the Mn--Zn ferrite has a cycle-work density roughly three orders of magnitude greater than those of Fe$_3$O$_4$ and MnFe$_2$O$_4$ throughout the plotted frequency range. The Fe$_3$O$_4$ and MnFe$_2$O$_4$ curves remain close and change relative order with frequency. At $\xi=0.1$ and $0.85$ [panel (b)], $W_{\mathrm{cyc}}$ increases with drive amplitude for all three ferrites. The Mn--Zn ferrite remains the largest at each frequency for both drive indices, while the ordering of Fe$_3$O$_4$ and MnFe$_2$O$_4$ depends on frequency and amplitude.

  The normalized out-of-phase susceptibility $\chi''/\chi_0$ and the cycle work $W_{\mathrm{cyc}}$ answer different questions. The ratio $\chi''/\chi_0$ is dimensionless and measures the out-of-phase magnetic response relative to that material's own zero-frequency in-phase susceptibility. By contrast, $W_{\mathrm{cyc}}$ is an absolute energy density transferred during one field cycle. In the linear limit to second order,
\[
W_{\mathrm{cyc}}^{(2)}=\pi\mu_0H_0^2\chi''=\pi\mu_0H_0^2\chi_0\left(\frac{\chi''}{\chi_0}\right),
\]
so a material can have a smaller value of $\chi''/\chi_0$ while still having a larger cycle work if its absolute susceptibility and/or imposed field amplitude is larger. Thus, Fig.~\ref{fig:family_chi}(b) compares the relative quadrature response of each material, whereas Fig.~\ref{fig:family_work} compares the absolute energy density transferred per cycle under the field condition assigned to each calculation.

\begin{figure}[t]
    \centering
    \includegraphics[width=\columnwidth]{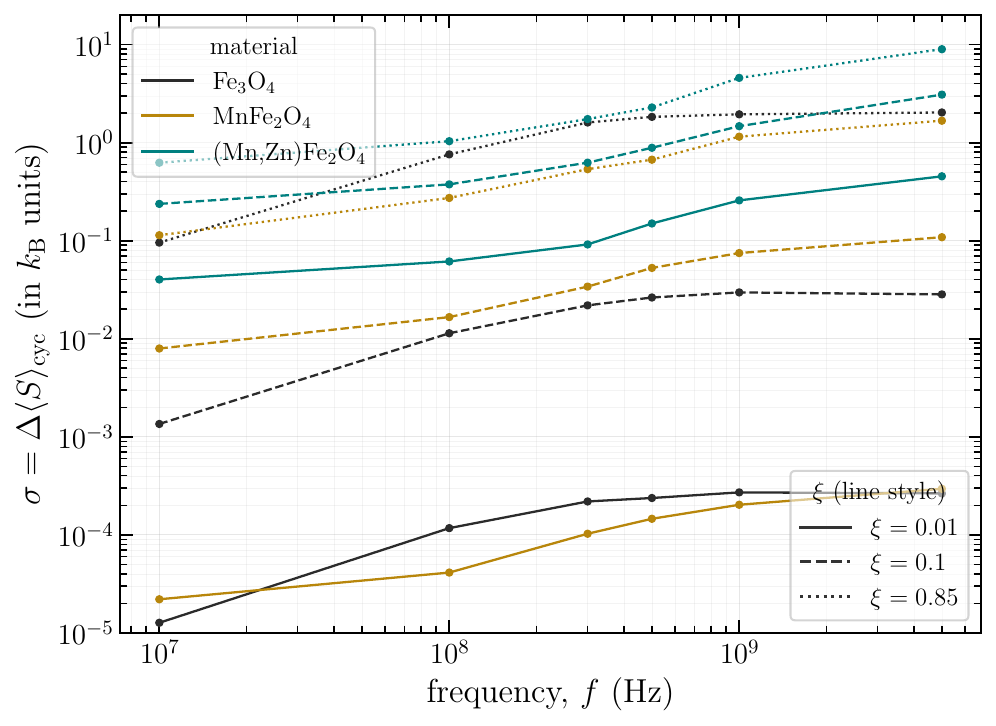}
    \caption{ Cycle-integrated entropy production versus cyclic drive frequency for the three ferrite families and the indicated drive-amplitude indices $\xi$.}
    \label{fig:family_entropy}
\end{figure}

The cycle-integrated entropy production in Fig.~\ref{fig:family_entropy} rises strongly with drive amplitude and is generally largest for \mnzn over the displayed conditions. Because it is integrated over one field period, it can be compared directly with the per-cycle work trends without introducing the additional factor associated with the number of cycles executed per unit time.

  Cycle work and entropy production are complementary but are not two definitions of the same quantity. $W_{\mathrm{cyc}}$ is the energy density transferred to the modeled material by the applied magnetic field during one cycle. The cycle-integrated entropy production, $\sigma_{\mathrm{cyc}}$, measures the irreversible redistribution of the electron, phonon, and magnon populations during that cycle. Because the modeled node is not maintained at a fixed temperature by an external reservoir, there is no single constant temperature that converts $\sigma_{\mathrm{cyc}}$ into $W_{\mathrm{cyc}}$ through a relation such as $W=T\sigma_{cyc}$. The two quantities can increase together when the drive produces larger departures from constrained equilibrium, but their numerical values have different physical meanings and different units.

\subsection{Nonlinear response}
\label{subsec:nonlinear}

\begin{figure}[t]
\centering
\includegraphics[width=\columnwidth]{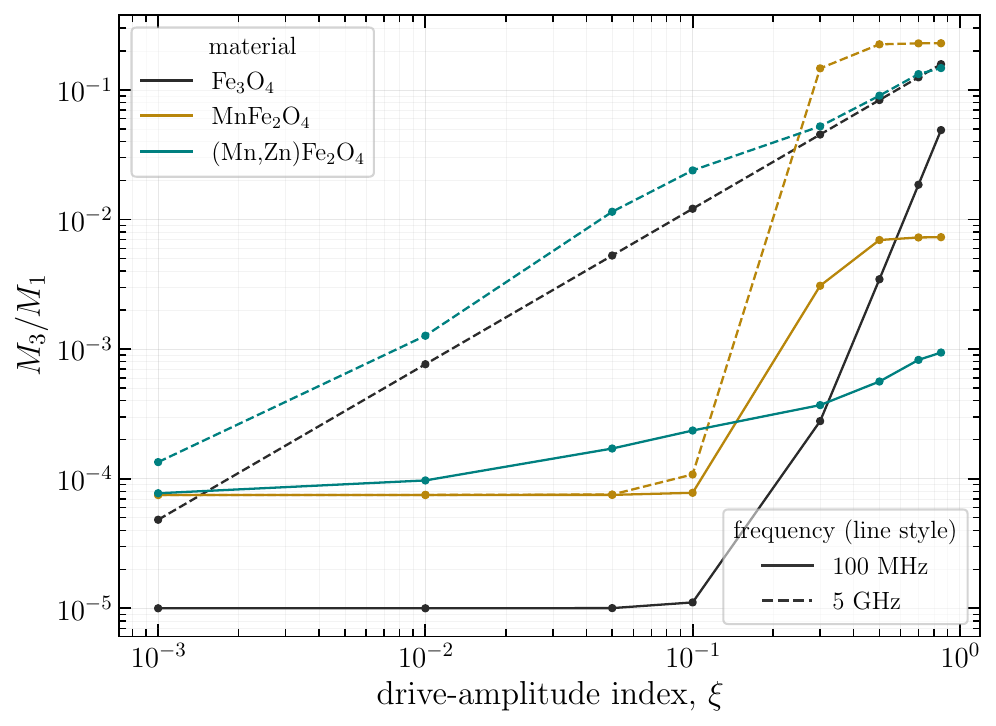}
\caption{Third-to-first harmonic ratio versus drive amplitude at 100 MHz and 5 GHz.}
\label{fig:family_harmonic}
\end{figure}

For a sinusoidal applied field with frequency $f$, the calculated magnetization can be decomposed into Fourier components at $f$, $2f$, $3f$, and higher integer multiples of the drive frequency. $M_1$ denotes the magnitude of the fundamental magnetization component at the drive frequency and $M_3$ denotes the magnitude of the third harmonic at $3f$. The small-signal regime is the low-amplitude range in which the magnetization is approximately linear in the applied field so that the response is dominated by $M_1$ and higher harmonics are negligible. Growth of $|M_3|/|M_1|$, therefore, measures departure from this approximately linear, nearly sinusoidal response.
The harmonic comparison in Fig.~\ref{fig:family_harmonic} shows how the response leaves the small-signal regime as the drive index $\xi$ increases. At low amplitude, the third harmonic is very small for all three families. As the drive grows, $|M_3|/|M_1|$ can increase by orders of magnitude, showing that a sinusoidal field no longer produces a nearly sinusoidal magnetization response.

When the comparison criterion is harmonic distortion measured by $|M_3|/|M_1|$, the ordering of the three materials depends on both drive frequency and drive amplitude. Around 100 MHz, the representative \fe curve develops the largest high-amplitude harmonic distortion, whereas \mnzn remains comparatively close to the fundamental response. At 5 GHz, the ordering changes. The Mn-containing systems rise much more strongly and \mnfe becomes the most distorted over part of the high-amplitude range. The nonlinear response,, thus, cannot be assigned as a fixed property of one chemistry without specifying frequency and field amplitude.

This result also helps separate waveform distortion from energy loss. A large third harmonic indicates departure from a first-harmonic constitutive response, but it does not represent an independent additive contribution to the cycle work $W_{\mathrm{cyc}}$ that can simply be added to $\Wcyc$. For the single-tone field used here, the loop work remains determined by the actual trajectory, while the harmonic ratio is best used as a measure of nonlinearity and the onset of waveform distortion.

\subsection{Subsystem temperatures}
\label{subsec:temperatures}

\begin{figure*}[t]
    \centering
    \subfloat[]{%
    \includegraphics[width=0.32\textwidth]{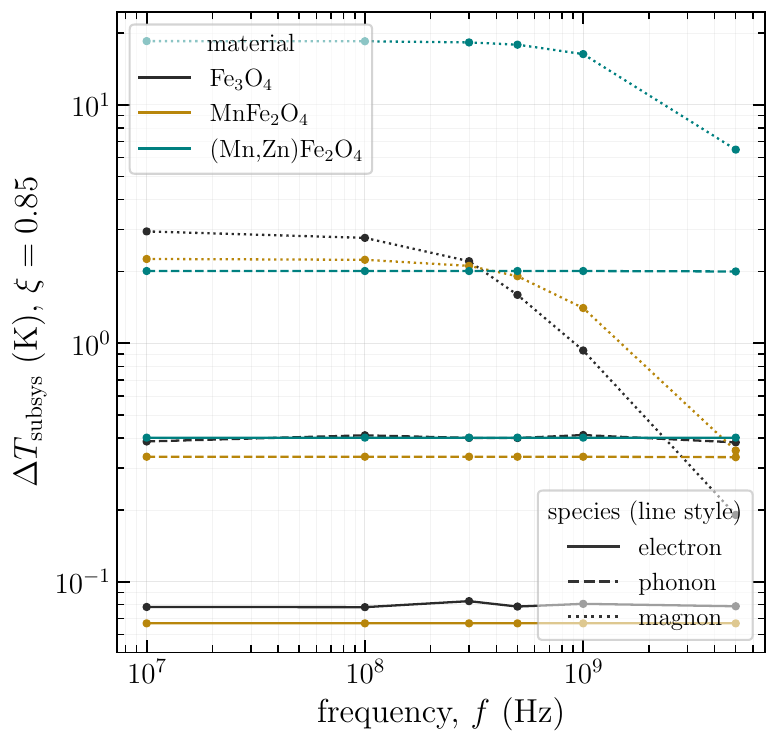}}
    \hfill
    \subfloat[]{%
    \includegraphics[width=0.32\textwidth]{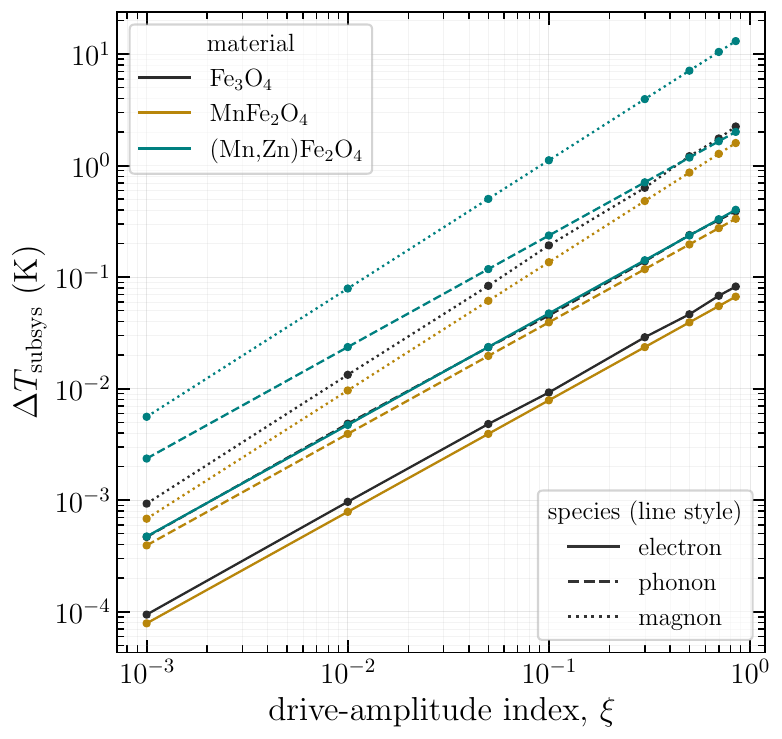}}
    \hfill
    \subfloat[]{%
    \includegraphics[width=0.32\textwidth]{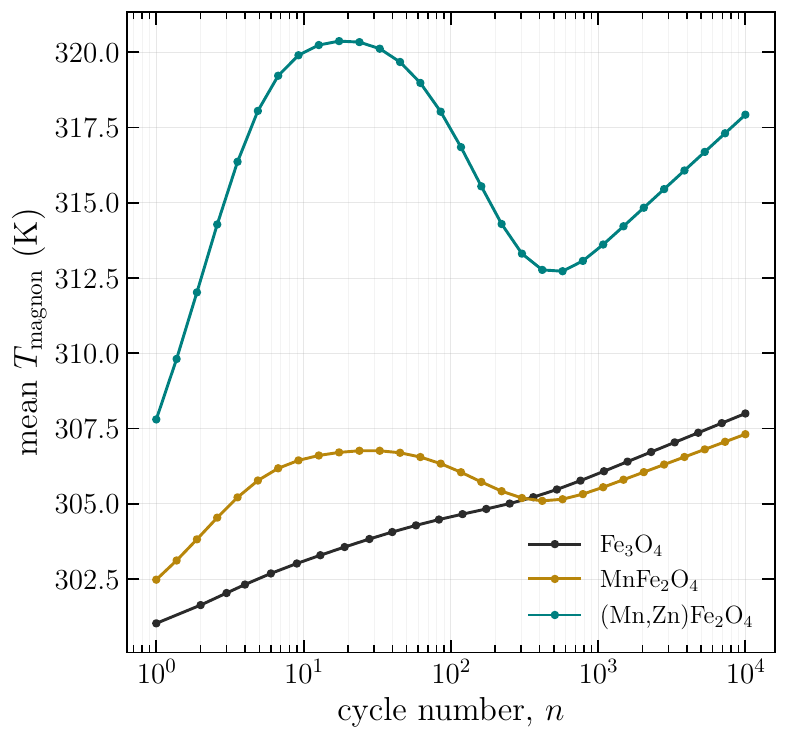}}
    \caption{ Subsystem-temperature response for the three ferrite families. (a) Peak-to-peak within-cycle temperature changes $\Delta T_{k,\mathrm{cyc}}$ for $k=e,p,m$ versus cyclic drive frequency $f$ at $\xi=0.85$. (b) Corresponding peak-to-peak temperature changes versus drive-amplitude index $\xi$ at $f=300$ MHz. (c) Mean magnon temperature versus cycle number at $f=300$ MHz and $\xi=0.85$.  }
    \label{fig:family_temperature}
\end{figure*}

 Fig.~\ref{fig:family_temperature} compares the temperature coordinates of the three excitation populations. For the populations $k\in\{e,p,m\}$, the peak-to-peak within-cycle temperature changes are defined as
\begin{equation}
    \Delta T_{k,\mathrm{cyc}}=\max_{t\in\mathrm{cycle}}T_k(t)-\min_{t\in\mathrm{cycle}}T_k(t).
\end{equation}
This quantity is a temperature difference, not an energy. Fig.~\ref{fig:family_temperature}(a) shows $\Delta T_{e,\mathrm{cyc}}$, $\Delta T_{p,\mathrm{cyc}}$, and $\Delta T_{m,\mathrm{cyc}}$ versus cyclic drive frequency at $\xi=0.85$. The magnon temperature change is the largest of the three populations because the applied field acts directly on the magnon energies. The Mn--Zn ferrite has the largest magnon temperature change over most of the plotted frequency range, while the corresponding electron and phonon changes remain much smaller. Toward the highest frequencies, the magnon temperature variation decreases as the magnon population becomes less able to follow the field during one period.

  Fig.~\ref{fig:family_temperature}(b) shows the amplitude dependence of the peak-to-peak temperature changes at $f=300$ MHz. The electron, phonon, and magnon values all increase with $\xi$, while maintaining the hierarchy $\Delta T_{m,\mathrm{cyc}}>\Delta T_{p,\mathrm{cyc}}>\Delta T_{e,\mathrm{cyc}}$ over the displayed range. Fig.~\ref{fig:family_temperature}(c) shows a different quantity: the mean magnon temperature over successive cycles at $f=300$ MHz and $\xi=0.85$. The nonmonotonic mean-$T_m$ behavior visible most strongly for the Mn--Zn ferrite does not imply that the total material node is cooling. It is instead consistent with competition between continuing magnetic work on the magnon population and redistribution of excitation energy from magnons to the electron and phonon populations. Because no external heat sink is present, the total excitation energy can continue to increase even during an interval in which the mean magnon temperature decreases.

 These curves show how the internal population-temperature coordinates respond to and redistribute the energy deposited by the applied field. They are not themselves direct measures of the transferred energy. Because an external thermal reservoir is not built into the model, the magnon temperatures cannot be associated with the actual operating temperatures of a ferrite core. A real device temperature would require geometry, thermal conductivity, interfaces, convection or coolant conditions, and all other loss mechanisms in addition to the longitudinal contribution predicted by the present model.

\subsection{Cation-configuration effects}
\label{subsec:config}

\begin{figure*}[t]
    \centering
    \subfloat[]{%
    \includegraphics[width=0.48\textwidth]{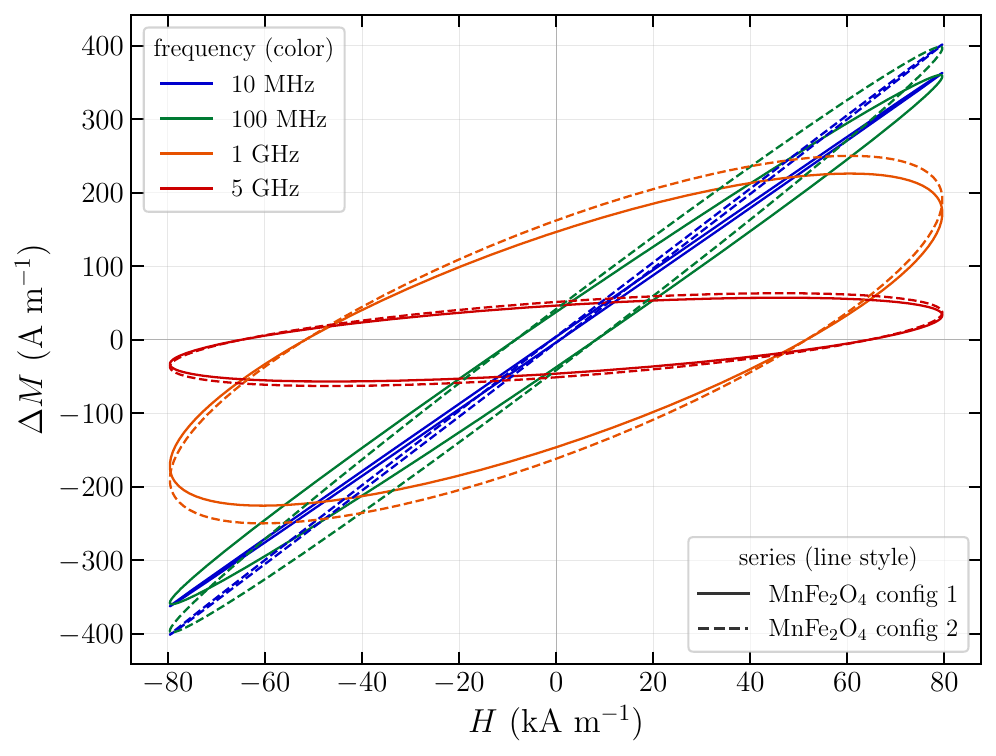}}
    \hfill
    \subfloat[]{%
    \includegraphics[width=0.48\textwidth]{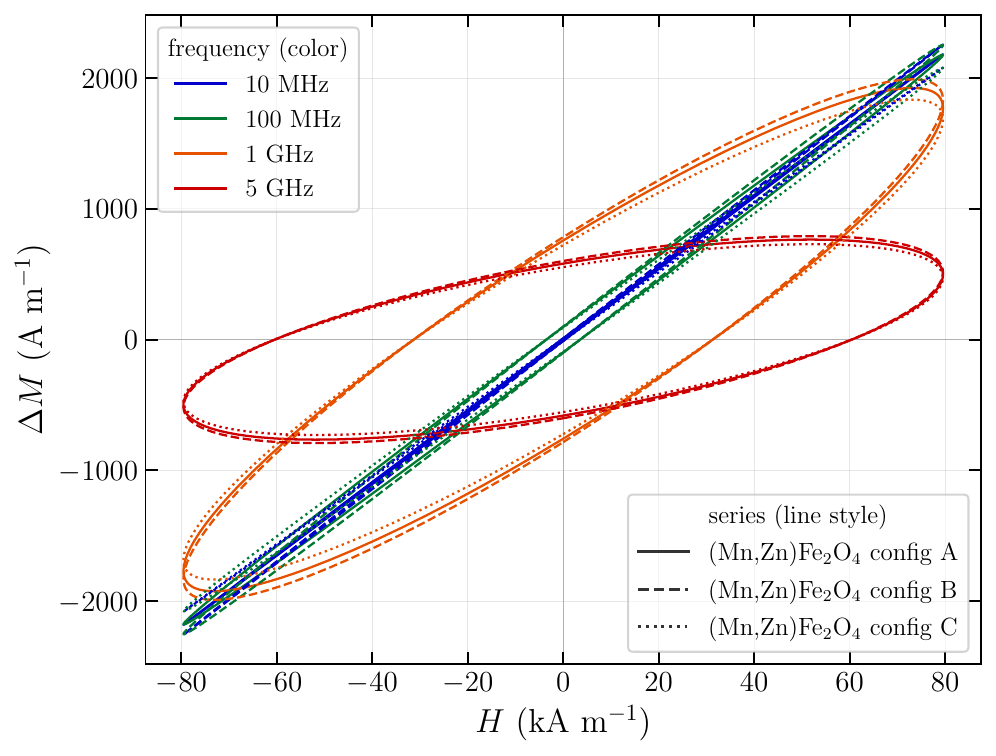}}
    \caption{ Configuration-resolved longitudinal $M$--$H$ trajectories: (a) MnFe$_2$O$_4$, configurations 1 and 2; (b) (Mn$_{0.5}$Zn$_{0.5}$)Fe$_2$O$_4$, configurations A--C.  }
    \label{fig:config_mh}
\end{figure*}

The configuration-resolved plots test whether the large family-level differences are accompanied by meaningful variations within one nominal chemistry. Fig.~\ref{fig:config_mh} compares the two \mnfe and three \mnzn arrangements. At 10 and 100 MHz, the alternative configurations of a given family lie close to one another. Their separation becomes easier to see at 1 and 5 GHz, where the response is already more sensitive to the details of the spectral distribution and to how readily the populations follow the field.

For \mnfe, configuration 2 generally produces the slightly larger loop opening and  peak longitudinal magnetization change at the more demanding operating points. For \mnzn, the three arrangements remain much closer to one another than the Mn--Zn family lies to \fe or \mnfe, but their high-frequency trajectories are not identical. Therefore, the relatively small differences within the two ferrite families suggest configuration (inversion and site substitution) is a small secondary effect in the present dataset relative to the larger influence of the family-level composition differences.

\begin{figure*}[t]
    \centering
    \subfloat[]{%
    \includegraphics[width=0.48\textwidth]{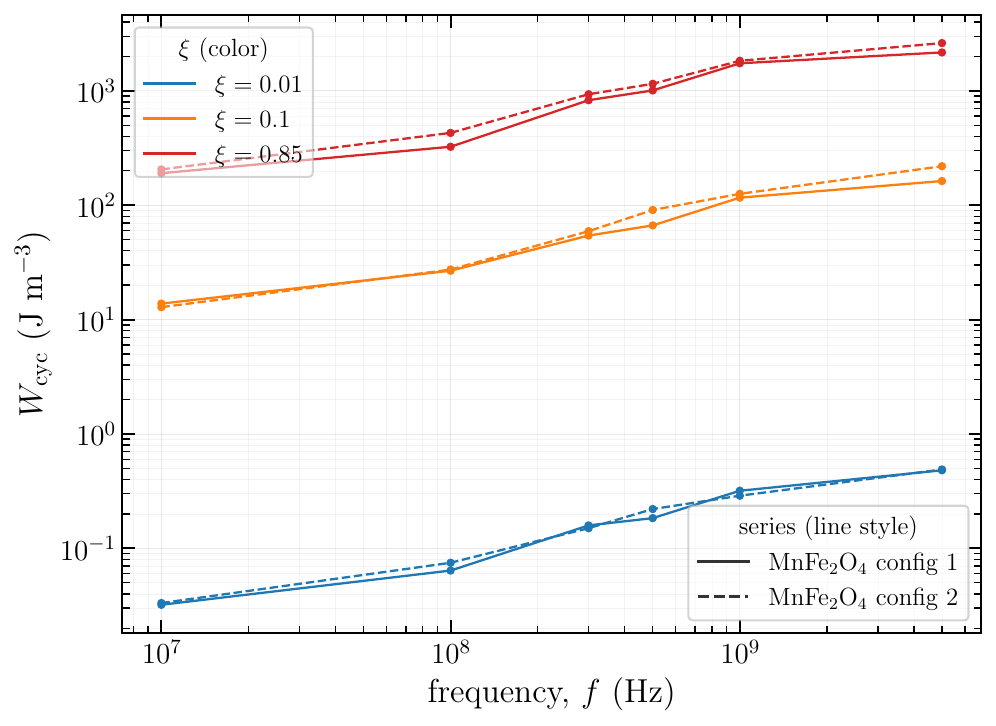}}
    \hfill
    \subfloat[]{%
    \includegraphics[width=0.48\textwidth]{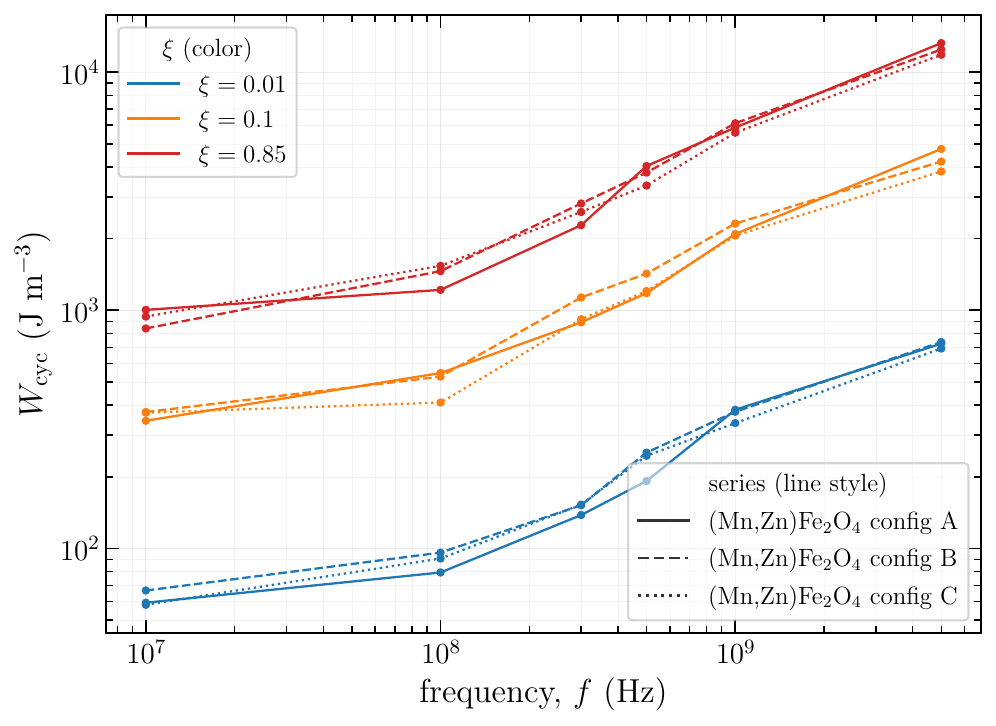}}
    \caption{ Configuration dependence of cycle work: (a) MnFe$_2$O$_4$, configurations 1 and 2; (b) (Mn$_{0.5}$Zn$_{0.5}$)Fe$_2$O$_4$, configurations A--C.  }
    \label{fig:config_work}
\end{figure*}

The same pattern appears in the work per cycle curves of Fig.~\ref{fig:config_work}. The two \mnfe configurations remain close but configuration 2 generally displays somewhat larger work per cycle. In the Mn--Zn set, the spread is modest at lower frequency and grows at high-frequencies. Configuration B is frequently on the high side of the work per cycle range, while configuration C is often lower. These differences are consistent with the configuration-dependent magnon and phonon spectra reported in Ref.~\cite{Dhariwal2026JCTC}.

\begin{figure}[t]
    \centering
    \includegraphics[width=\columnwidth]{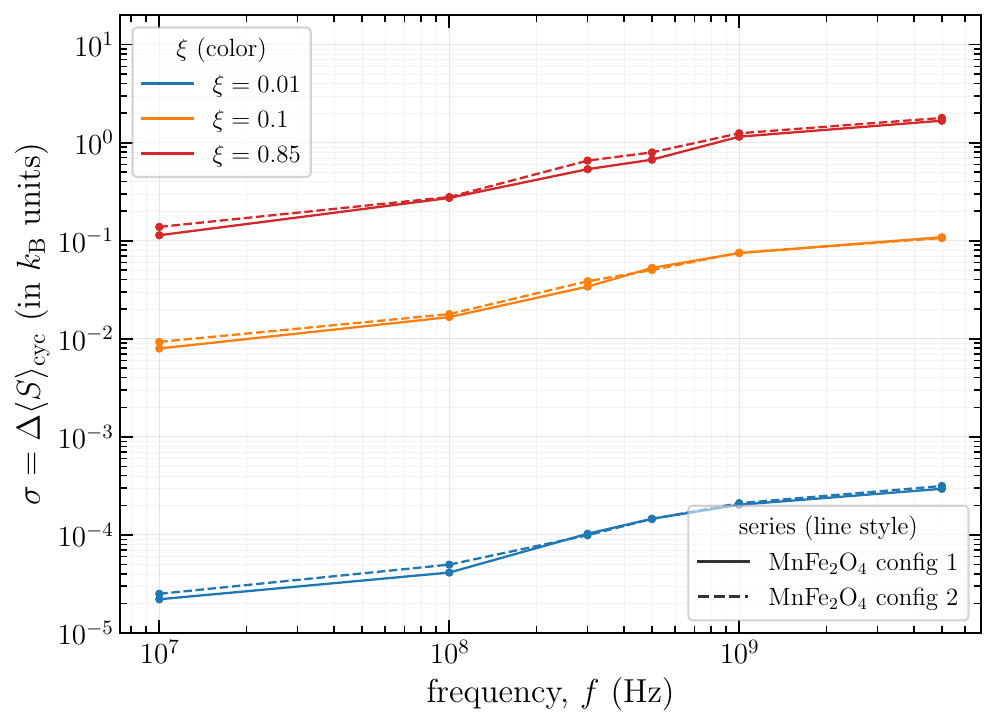}
    \caption{Configuration dependence of cycle-integrated entropy production for MnFe$_2$O$_4$.  }
    \label{fig:config_entropy_mnfe}
\end{figure}

\begin{figure}[t]
    \centering
    \includegraphics[width=\columnwidth]{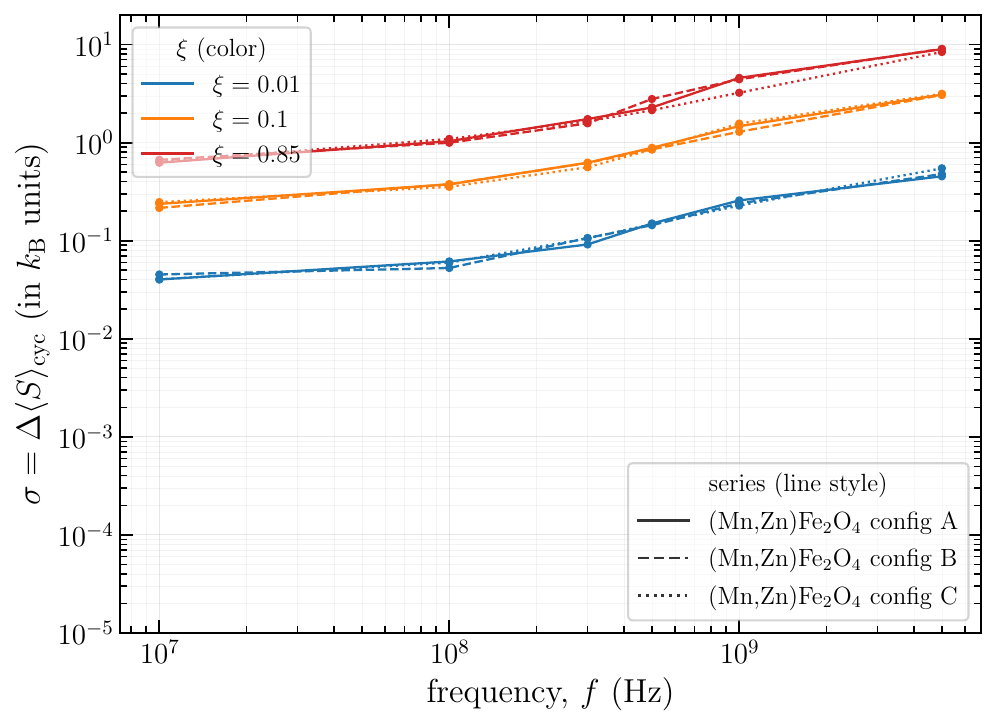}
    \caption{Configuration dependence of cycle-integrated entropy production for (Mn$_{0.5}$Zn$_{0.5}$)Fe$_2$O$_4$.   }
    \label{fig:config_entropy_mnzn}
\end{figure}

 The configuration-resolved cycle-integrated entropy results are shown in Figs.~\ref{fig:config_entropy_mnfe} and~\ref{fig:config_entropy_mnzn}. For MnFe$_2$O$_4$, configuration 2 generally gives the larger $\sigma_{\mathrm{cyc}}$, with the separation becoming more visible as $\xi$ increases and toward the upper end of the plotted frequency range. In the Mn--Zn set, the three configurations remain close at smaller $\xi$ and lower frequency, while their separation is most visible at the largest plotted drive index, $\xi=0.85$, in the upper-frequency portion of the sweep. The comparison is, therefore, made using the entropy generated per cycle rather than an additional entropy-production-rate plot.

\begin{figure*}[t]
    \centering
    \subfloat[]{%
    \includegraphics[width=0.32\textwidth]{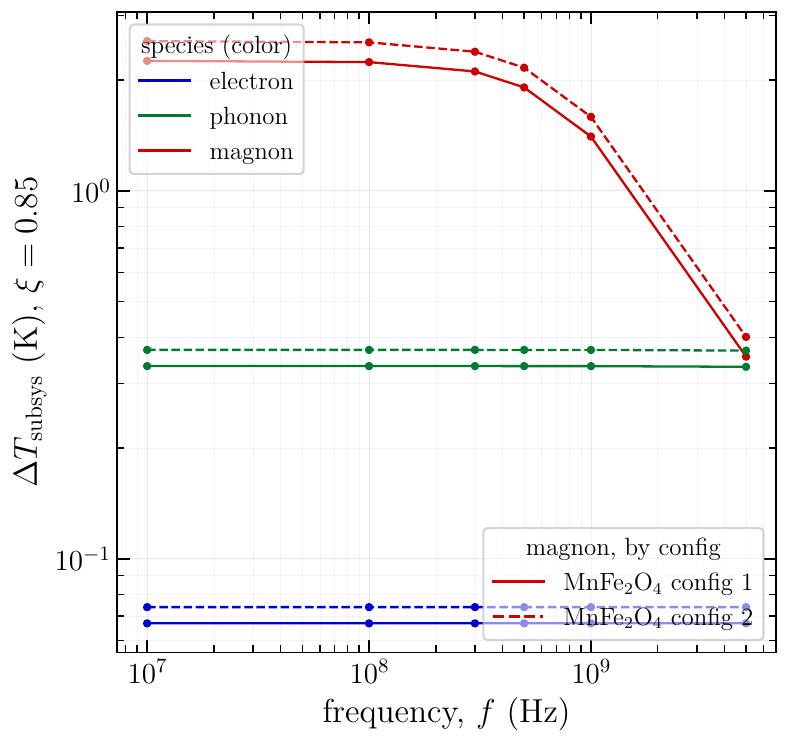}}
    \hfill
    \subfloat[]{%
    \includegraphics[width=0.32\textwidth]{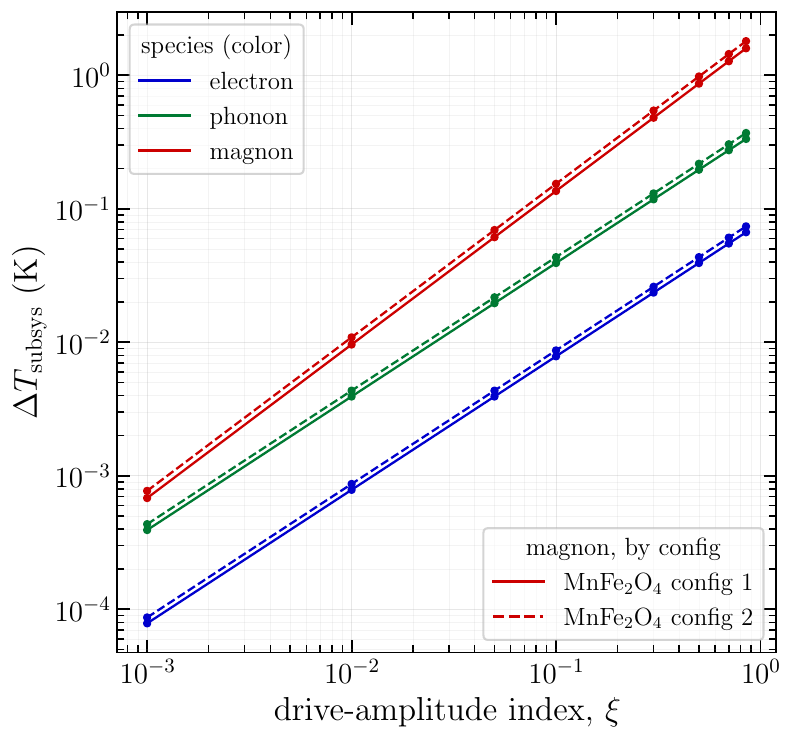}}
    \hfill
    \subfloat[]{%
    \includegraphics[width=0.32\textwidth]{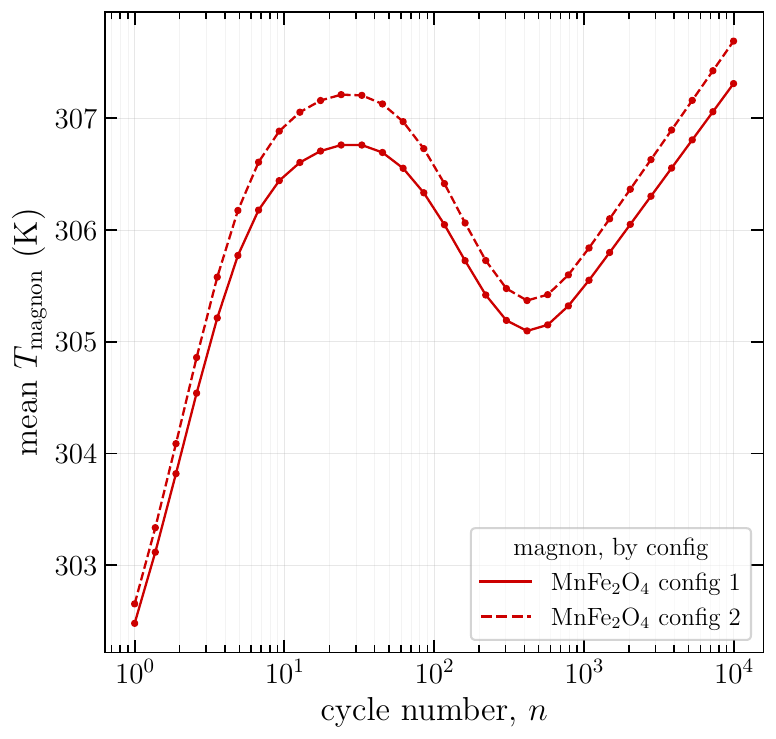}}
    \caption{Configuration dependence of the population-temperature response for MnFe$_2$O$_4$. (a) Peak-to-peak within-cycle electron, phonon, and magnon temperature changes versus cyclic drive frequency at $\xi=0.85$. (b) Corresponding peak-to-peak temperature changes versus drive-amplitude index $\xi$. (c) Mean magnon temperature versus cycle number. Species are identified by color and configurations 1 and 2 by line style.}
    \label{fig:config_temperature_mnfe}
\end{figure*}

\begin{figure*}[t]
    \centering
    \subfloat[]{%
    \includegraphics[width=0.32\textwidth]{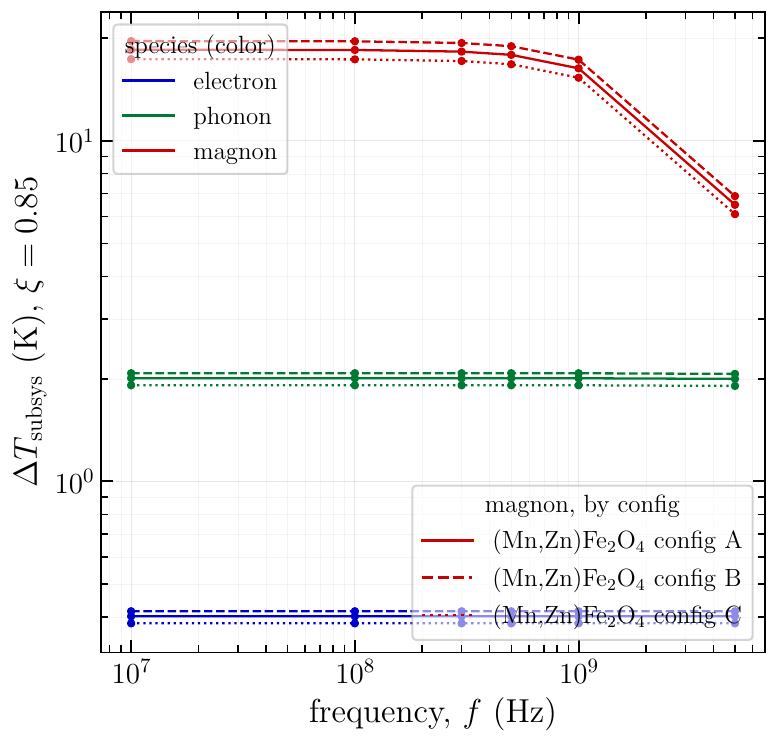}}
    \hfill
    \subfloat[]{%
    \includegraphics[width=0.32\textwidth]{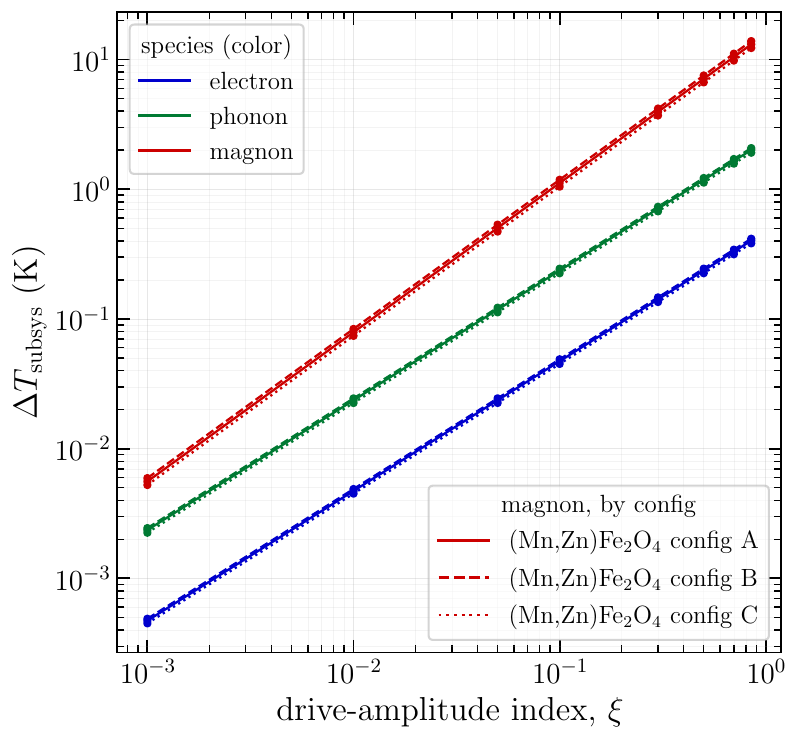}}
    \hfill
    \subfloat[]{%
    \includegraphics[width=0.32\textwidth]{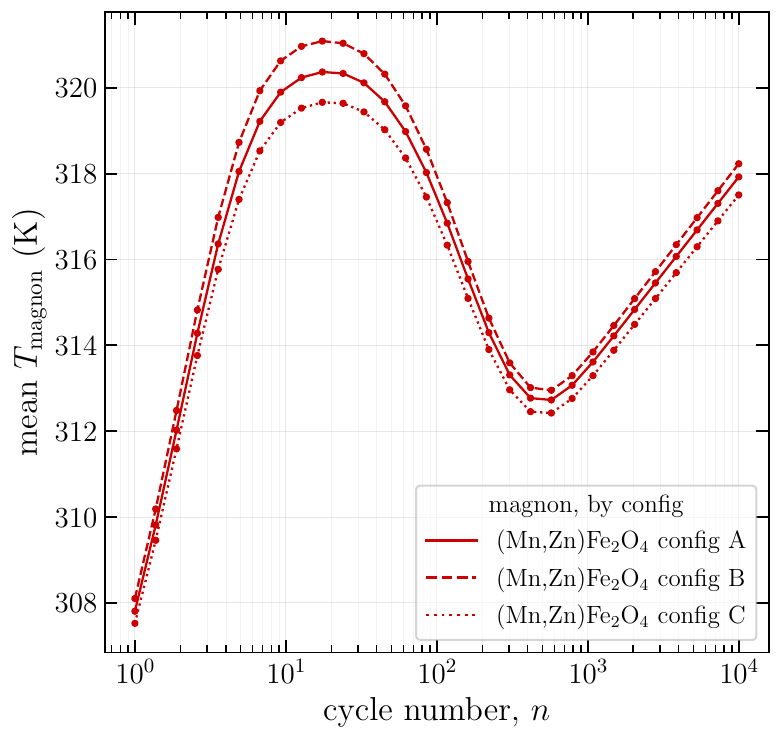}}
    \caption{Configuration dependence of the population-temperature response for (Mn$_{0.5}$Zn$_{0.5}$)Fe$_2$O$_4$. (a) Peak-to-peak within-cycle electron, phonon, and magnon temperature changes versus cyclic drive frequency at $\xi=0.85$. (b) Corresponding peak-to-peak temperature changes versus drive-amplitude index $\xi$. (c) Mean magnon temperature versus cycle number. Species are identified by color and configurations A--C by line style.}
    \label{fig:config_temperature_mnzn}
\end{figure*}

     Figures~\ref{fig:config_temperature_mnfe} and~\ref{fig:config_temperature_mnzn} compare the population-temperature response among cation configurations. In panels (a) and (b), the magnon population has the largest peak-to-peak temperature change, the phonon population is intermediate, and the electron population has the smallest change. This ordering follows the model coupling: the magnetic field acts directly on the magnon energies, while the electron and phonon populations respond through the shared energy constraint.

For MnFe$_2$O$_4$, configuration 2 gives slightly larger peak-to-peak temperature changes than configuration 1 over most of the displayed conditions. In panel (c), the mean magnon temperatures of both configurations first rise, then decrease over an intermediate cycle-number range, and then rise again. Configuration 2 remains slightly above configuration 1. For the Mn--Zn ferrite, configuration B generally has the highest mean magnon temperature at a given cycle number, configuration C the lowest, and configuration A lies between them over most of the displayed cycle-number range. The nonmonotonic mean-$T_m$ trajectories in panels (c) are consistent with competition between direct magnetic work on the magnon population and internal transfer of excitation energy from magnons to electrons and phonons. A temporary decrease in mean $T_m$, thus, represents redistribution within the isolated three-population node and does not imply a decrease in the node's total excitation energy. Continued positive cycle work without a heat interaction with the surroundings produces the later increase in the mean temperatures.

\begin{figure}[t]
    \centering
    \includegraphics[width=\columnwidth]{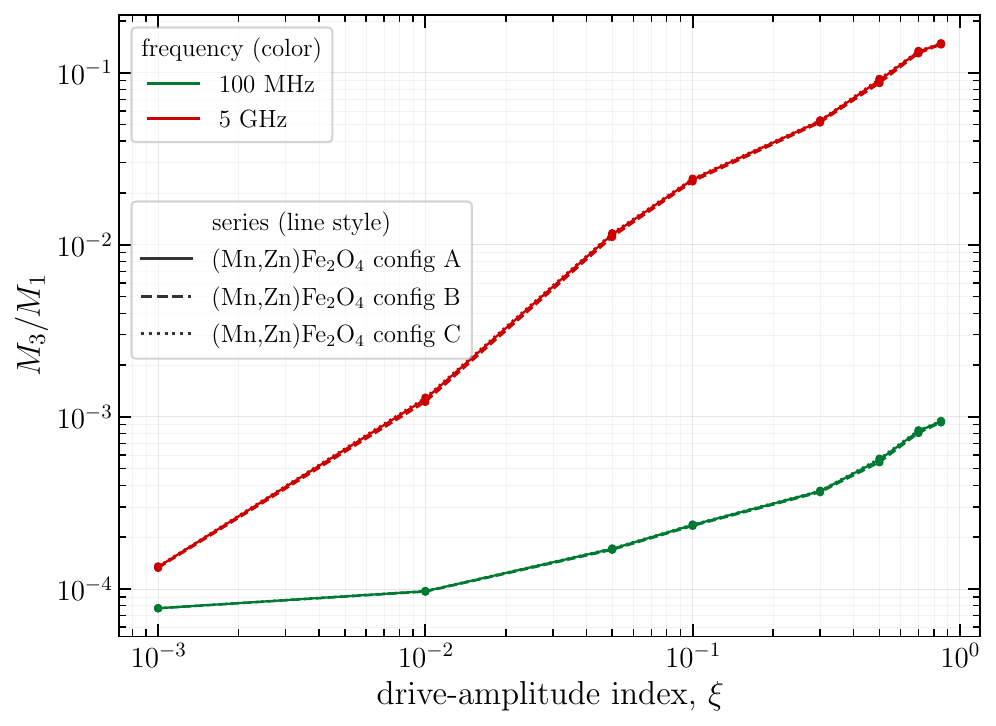}
    \caption{Mn--Zn configuration dependence of the third-harmonic ratio.}
    \label{fig:mnzn_harm_config}
\end{figure}

The configuration dependence of $|M_3|/|M_1|$ is much weaker than the configuration dependence of cycle work and subsystem-temperature response. In Fig.~\ref{fig:mnzn_harm_config}, the A--C curves nearly overlap at both 100 MHz and 5 GHz even though the same configurations show visible differences in work and temperature. This is a useful reminder that cation rearrangement need not affect every observable with the same sensitivity. A change can be energetically significant without producing a comparably large change in the normalized harmonic metric.

These configuration trends are consistent with the first-principles spectral results reported in Ref.~\cite{Dhariwal2026JCTC}, particularly the configuration-dependent exchange constants in Table~3 of that work, the MnFe$_2$O$_4$ magnon spectra in Fig.~11, the Mn--Zn magnon spectra in Fig.~12, and the phonon-DOS comparison in Fig.~13. Moving Mn between A and B sites changes the dominant exchange paths, changes how the magnon density of states is distributed over excitation energy, and modifies low- and mid-energy phonon structure~\cite{Dhariwal2026JCTC}. Experimental studies of Zn-substituted manganese and Mn--Zn ferrites likewise show that Zn content, cation distribution, solid-state synthesis conditions, and the resulting microstructure can move magnetic relaxation frequencies and alter loss~\cite{Praveena2016,Zapata2013}.  The present calculations, thus, show that cation arrangement is a secondary source of variation, after composition, in the frequency dependence of the modeled longitudinal response.

\subsection{Engineering interpretation}
\label{subsec:screening}

Table~\ref{tab:screening} summarizes trends from the aforementioned figures of relaxation in an alternating magnetic field. Since the energy eigenstructures employed for electrons, phonons, and magnons omit consideration of  features like domain walls, transverse precession, or macroscopic reversal and the particular form of the SEAQT equation of motion neglects specimen-scale electromagnetic diffusion and heat-exchange with surroundings, the table summary does not necessarily reflect the performance of an actual magnetic ferrite core. Nevertheless, it is useful for understanding the frequency- and amplitude-dependence of losses arising from longitudinal magnon relaxation.

\begin{table*}[t]
    \caption{Comparison of specific observables from the modeled longitudinal relaxation response; each entry refers to the plotted frequency and drive-amplitude ranges and does not represent total core performance. \\ }
    \label{tab:screening}
    \centering
    \begingroup
    \setlength{\tabcolsep}{9pt}
    \renewcommand{\arraystretch}{1.08}
    \begin{tabular}{@{}llll@{}}
    \toprule
    \parbox[t]{0.13\textwidth}{Metric} &
    \parbox[t]{0.18\textwidth}{\fe} &
    \parbox[t]{0.18\textwidth}{\mnfe} &
    \parbox[t]{0.21\textwidth}{\mnzn} \\
    \midrule
    \parbox[t]{0.13\textwidth}{Peak longitudinal magnetization change $|\Delta M|_{\mathrm{peak}}$} &
    \parbox[t]{0.18\textwidth}{Begins its strong decrease near $10^8$ Hz and becomes smaller than the \mnfe value at high frequency} &
    \parbox[t]{0.18\textwidth}{Similar to the \fe value at low frequency but remains larger at high frequency} &
    \parbox[t]{0.21\textwidth}{Approximately an order of magnitude larger at the low-frequency end and remains largest over the plotted range} \\[2pt]
    \midrule
    \parbox[t]{0.13\textwidth}{Normalized out-of-phase susceptibility $\chi''/\chi_0$} &
    \parbox[t]{0.18\textwidth}{Becomes appreciable at lower angular frequency than the Mn--Zn response and remains appreciable over a broad interval; its largest plotted value is below that of \mnfe} &
    \parbox[t]{0.18\textwidth}{Becomes appreciable at relatively low angular frequency, persists over a broad interval, and attains the largest plotted normalized value} &
    \parbox[t]{0.21\textwidth}{Is displaced toward higher angular frequency and attains a smaller largest plotted normalized value than \mnfe} \\[2pt]
    \midrule
    \parbox[t]{0.13\textwidth}{Cycle-work density $W_{\mathrm{cyc}}$} &
    \parbox[t]{0.18\textwidth}{Remains below the Mn--Zn value; its ordering relative to \mnfe depends on frequency and $\xi$} &
    \parbox[t]{0.18\textwidth}{Is comparable to the \fe value, with frequency- and amplitude-dependent crossovers} &
    \parbox[t]{0.21\textwidth}{Is the largest throughout the plotted frequency range for $\xi=0.01$, 0.1, and 0.85} \\[2pt]
    \midrule
    \parbox[t]{0.13\textwidth}{Cycle-integrated entropy production $\sigma_{\mathrm{cyc}}$} &
    \parbox[t]{0.18\textwidth}{Is below the Mn--Zn value over most plotted frequencies and amplitudes} &
    \parbox[t]{0.18\textwidth}{Is comparable to the \fe value and crosses it as frequency or $\xi$ changes} &
    \parbox[t]{0.21\textwidth}{Is generally the largest over the displayed frequency and amplitude conditions} \\[2pt]
    \midrule
    \parbox[t]{0.13\textwidth}{Harmonic distortion $|M_3|/|M_1|$} &
    \parbox[t]{0.18\textwidth}{Is the largest of the three families at the upper sampled $\xi$ values around 100 MHz} &
    \parbox[t]{0.18\textwidth}{Is the largest over the upper part of the sampled amplitude range at 5 GHz} &
    \parbox[t]{0.21\textwidth}{Is the smallest at 100 MHz but increases substantially over the sampled amplitudes in the GHz range} \\[2pt]
    \midrule
    \parbox[t]{0.13\textwidth}{Peak-to-peak magnon temperature change $\Delta T_{m,\mathrm{cyc}}$} &
    \parbox[t]{0.18\textwidth}{Remains below the Mn--Zn value over most of the plotted frequency sweep at $\xi=0.85$} &
    \parbox[t]{0.18\textwidth}{Remains below the Mn--Zn value over most of the plotted frequency sweep at $\xi=0.85$} &
    \parbox[t]{0.21\textwidth}{Is the largest over most of the plotted frequency sweep at $\xi=0.85$} \\
    \bottomrule
    \end{tabular}
    \endgroup
\end{table*}

Several engineering implications follow from the calculated observables. First, $\chi''/\chi_0$ and $W_{\mathrm{cyc}}$ should not be interpreted as interchangeable measures: $\chi''/\chi_0$ gives the out-of-phase response relative to each material's own zero-frequency susceptibility, whereas $W_{\mathrm{cyc}}$ gives the absolute energy density transferred during one cycle under the imposed field condition. Second, the frequency at which $\chi'$ begins to decrease and the frequency range over which $\chi''$ is appreciable cannot be inferred from $\tau_m$ alone. Eq.~\eqref{eq:chi} also contains the electron and phonon relaxation parmeters and energy-weighted contributions from the electron, phonon, and magnon densities of states. Third, harmonic distortion is not fixed by composition alone because $|M_3|/|M_1|$ changes with both drive frequency and amplitude. Finally, different cation arrangements produce measurable changes in the $M$--$H$ response, cycle work, entropy production, and population-temperature response, but these within-composition differences are smaller than the family-to-family differences in the present data set. The present calculations, therefore, establish sensitivity to the cation configuration but not an optimization of cation arrangement.


\section{Limitations and validation}
\label{sec:limitations}

The first uncertainty lies in the excitation spectra. The electron, phonon, and magnon DOS were calculated in a common cubic-spinel reference using harmonic phonon and spin-wave descriptions~\cite{Dhariwal2026JCTC}. This controlled reference is valuable for comparison, but it does not include every finite-temperature renormalization, anharmonic process, critical magnetic fluctuation, defect state, or processing-induced phase. Close to a Curie point or in a strongly nonstoichiometric specimen, the input eigenstructure itself may need to be updated before the relaxation result can be quantitative.

\begin{figure*}[t]
    \centering
    \subfloat[]{%
    \includegraphics[width=0.48\textwidth]{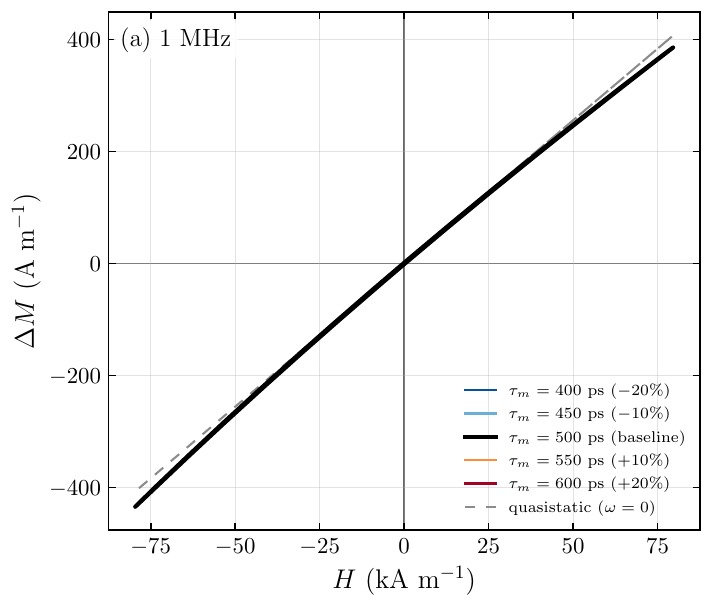}}
    \hfill
    \subfloat[]{%
    \includegraphics[width=0.48\textwidth]{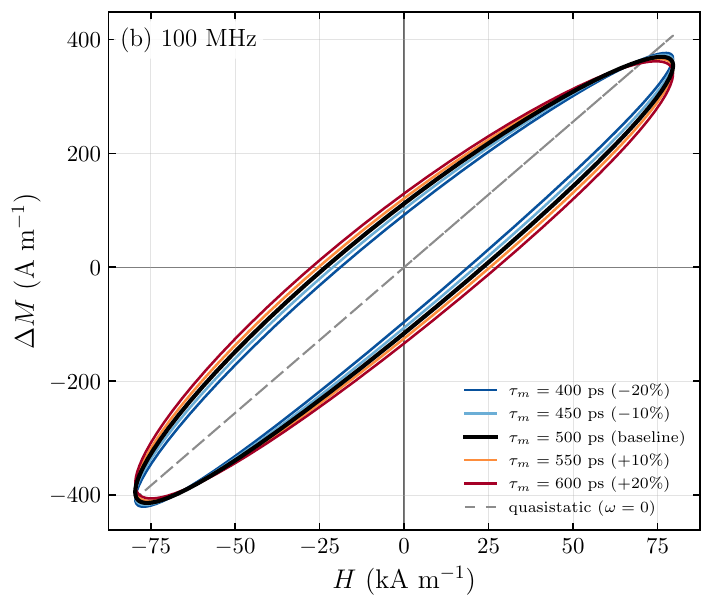}}
    \par\medskip
    \subfloat[]{%
    \includegraphics[width=0.48\textwidth]{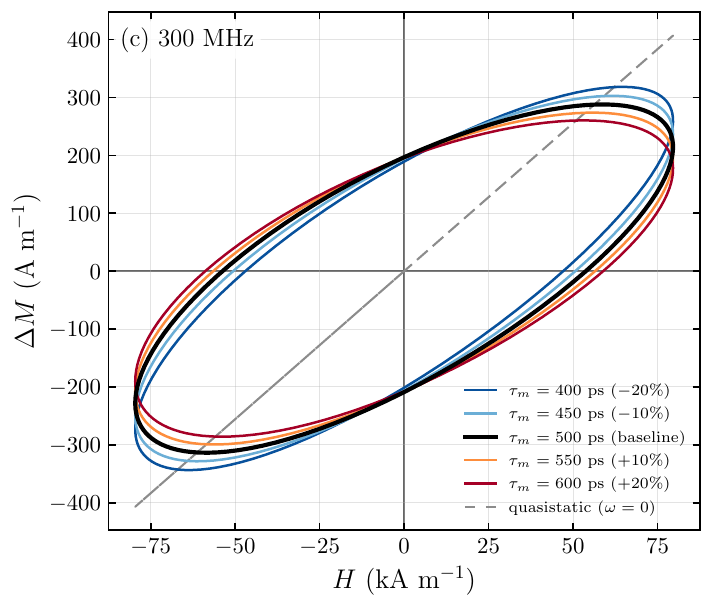}}
    \hfill
    \subfloat[]{%
    \includegraphics[width=0.48\textwidth]{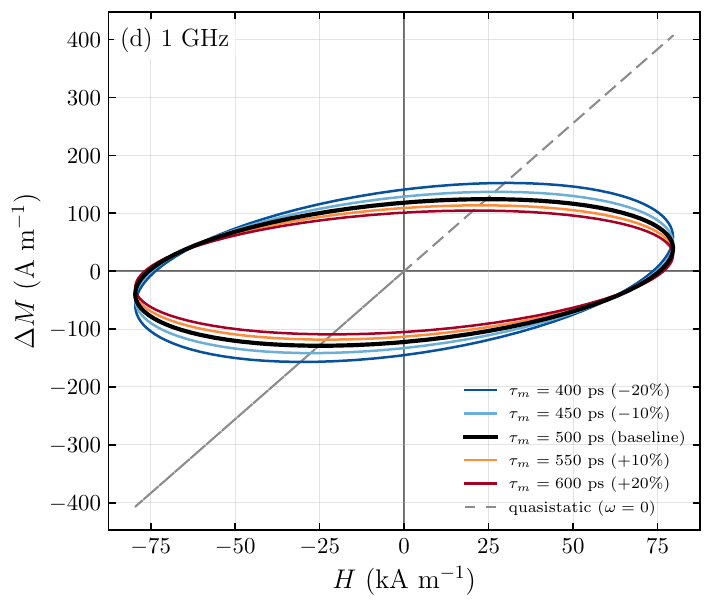}}
    \caption{Sensitivity of the longitudinal Fe$_3$O$_4$ $M$--$H$ trajectory to the prescribed magnon-population relaxation parameter $\tau_m$. The 500-ps baseline is compared with variations of $-20\%$, $-10\%$, $+10\%$, and $+20\%$ at (a) 1 MHz, (b) 100 MHz, (c) 300 MHz, and (d) 1 GHz. The quasistatic trajectory is shown for reference.}
    \label{fig:tau_sensitivity_mh}
\end{figure*}

\begin{figure*}[t]
    \centering
    \subfloat[]{%
    \includegraphics[width=0.48\textwidth]{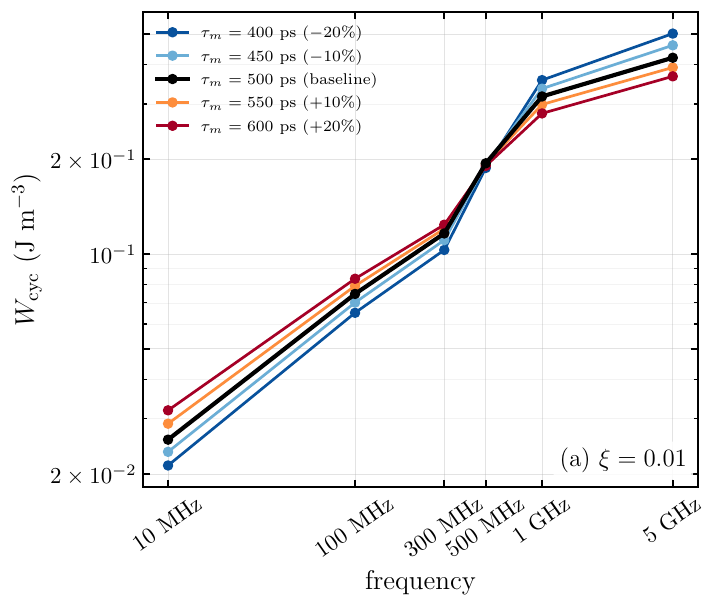}}
    \hfill
    \subfloat[]{%
    \includegraphics[width=0.48\textwidth]{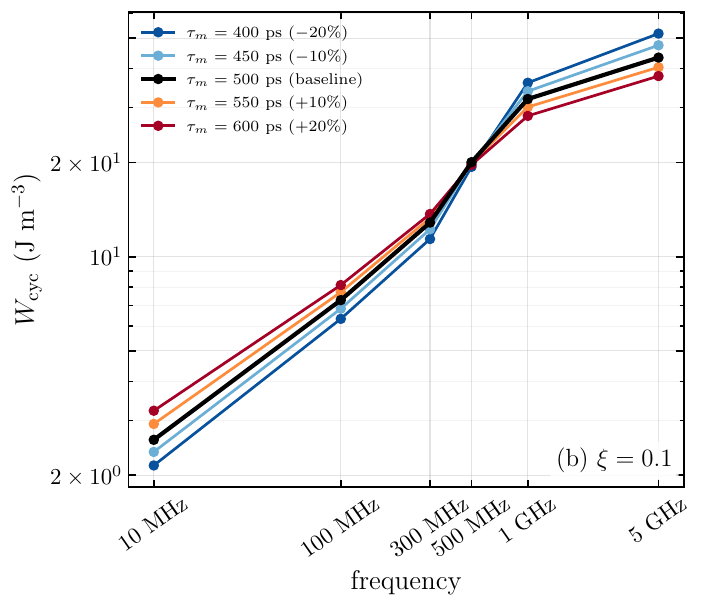}}
    \par\medskip
    \subfloat[]{%
    \includegraphics[width=0.48\textwidth]{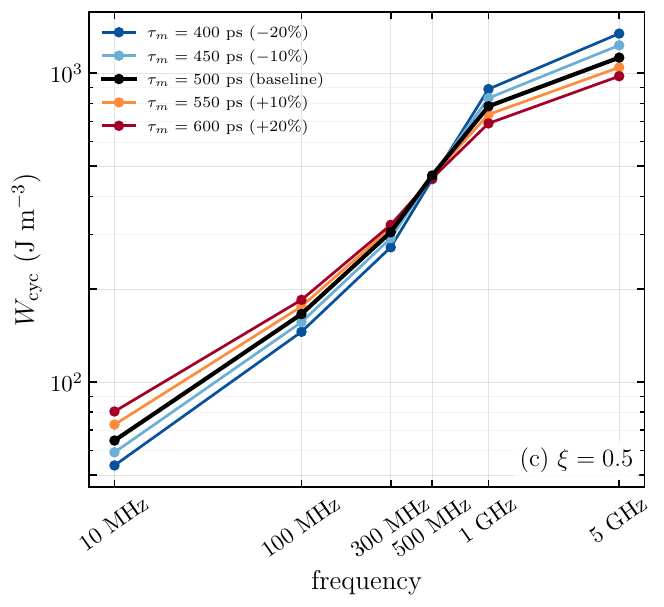}}
    \hfill
    \subfloat{%
    \includegraphics[width=0.48\textwidth]{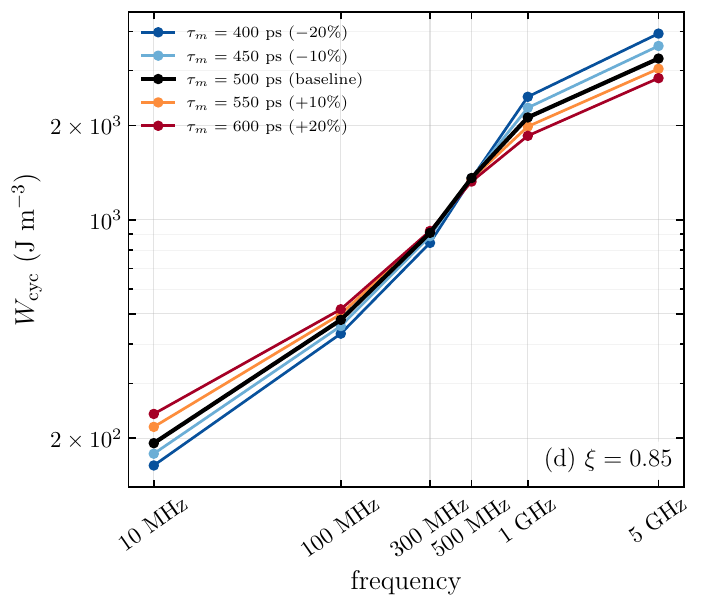}}
    \caption{Sensitivity of the Fe$_3$O$_4$ cycle-work density $W_{\mathrm{cyc}}$ to $\tau_m$ over the sampled frequencies at (a) $\xi=0.01$, (b) $\xi=0.1$, (c) $\xi=0.5$, and (d) $\xi=0.85$. The five curves span $\tau_m=400$--600 ps about the 500-ps baseline.}
\label{fig:tau_sensitivity_work}
\end{figure*}

The second uncertainty is dynamic. Section~\ref{subsec:kinetic_provenance} explains the physical basis for the selected hierarchy of 0.05 ps, 3 ps, and sub-nanosecond magnetic times. The evidence is strongest for the orders of magnitude, not for a unique mapping of each experimental decay constant onto the SEAQT dynamic parameters. The common $\tau_e$ and $\tau_p$ are controlled comparison assumptions, while the material-specific $\tau_m$ values are literature-bounded estimates. To quantify the consequence of this uncertainty, Figs.~\ref{fig:tau_sensitivity_mh} and~\ref{fig:tau_sensitivity_work} vary the Fe$_3$O$_4$ value of $\tau_m$ from 400 to 600 ps, corresponding to $\pm20\%$ about the 500-ps baseline, while holding the excitation spectra and the other model inputs fixed.

At 1 MHz, the five $M$--$H$ trajectories are visually indistinguishable and remain close to the quasistatic line, showing that a $\pm20\%$ change in $\tau_m$ has negligible influence when the drive period is long relative to the magnetic-population relaxation parameter. Separation develops at 100 and 300 MHz and is pronounced by 1 GHz. The associated work curves reveal a crossover rather than a uniform sensitivity: below the crossover, increasing $\tau_m$ increases $W_{\mathrm{cyc}}$, whereas above it the ordering reverses and the shorter relaxation parameters give the larger work. The curves nearly intersect between 300 and 500 MHz for every plotted $\xi$, and the same change of ordering is visible in the loop openings.

This behavior is physically consistent with moving the characteristic magnetic-relaxation spectrum relative to a fixed drive frequency. On the low-frequency side of a relaxation feature, a larger $\tau_m$ increases the phase lag accumulated during a cycle and, therefore, increases the loop area. On the high-frequency side, a shorter $\tau_m$ allows the magnetic population to follow more of the imposed variation, which increases the absolute out-of-phase response and cycle work. Although the coupled SEAQT response is not governed by a single Debye time, the crossover is the expected qualitative consequence of shifting a dissipative relaxation feature through the observation window. The sensitivity is appreciable away from the crossover but remains bounded: varying $\tau_m$ by $\pm20\%$ changes the quantitative work values without changing their order of magnitude or the principal family-level conclusions. Quantitative predictions should, nevertheless, propagate uncertainty in all three species relaxation parameters and allow them to vary with temperature, cation distribution, defect content, and processing when suitable measurements become available.

A further limitation is mechanism completeness. Measured ferrite loss can contain domain-wall motion, magnetization rotation, resonance, spin damping, and eddy-current contributions~\cite{Visser1984,Lebourgeois1996,Beatrice2006,Tsutaoka1999,Dobak2022,Wu2024}. Those mechanisms vary strongly with grain size, dc bias, processing, and geometry. The present calculation is most naturally tested in regimes where the longitudinal intra-domain contribution can be isolated or where the omitted mechanisms can be estimated independently. Bias-dependent MHz measurements are particularly useful because they show how strongly finished-core loss changes when the magnetic state and loop shape are altered~\cite{Sanusi2023}.

Finally, the model has no external heat sink. Actually, this is a simplifying assumption that could be relaxed rather than an inherent limitation of the SEAQT approach. Nevertheless, the absence of a heat interaction with the surroundings implies the subsystem temperatures describe internal redistribution of absorbed energy, not a final device temperature. A practical thermal prediction requires conduction through the core and interfaces, winding losses, packaging, and an environmental heat interaction in addition to the present magnetic source term.

These limitations suggest a direct validation path. Broadband complex permeability can test the predicted dispersion; harmonic analysis under increasing sinusoidal field can test the nonlinear onset; time-resolved probes can constrain the species-level dynamic scales; and total core-loss measurements on well-characterized specimens can establish how large the isolated longitudinal contribution is relative to domain and electromagnetic losses. The goal of such a program should not be to force measured total loss to equal Eq.~\eqref{eq:work}. It should be to determine when the quasiparticle contribution is significant and how reliably the spectrum-based model predicts its material dependence.

\section{Conclusions}
\label{sec:conclusions}

A common field-driven SEAQT formalism has been applied to the electron, phonon, and magnon spectra of \fe, \mnfe, and \mnzn. The comparison uses a fixed electronic relaxation parameter of 0.05 ps and a fixed phonon relaxation parameter of 3 ps for all ferrite types together with effective longitudinal magnon-population relaxation parameters of 500, 200, and 85 ps for \fe, \mnfe, and \mnzn, respectively. The electron and phonon values are deliberately held common to preserve a controlled comparison while the DOS retain their material specificity. Ultrafast magnetite measurements support a tens-of-femtoseconds electronic scale and a picosecond lattice scale, whereas magnetic measurements support a broader sub-nanosecond range for the slower magnetic sector. None of these inputs is fitted to the calculated work or susceptibility curves.

Across the plotted frequency ranges, \mnzn exhibits the largest change in longitudinal magnetization, the largest high-frequency retention, the largest cycle work, and the largest magnon-temperature change of the three ferrite families. Its normalized $\chi''$ peak is, nevertheless, broader and lower than the representative \mnfe peak. 

The comparison between \fe and \mnfe is more dependent upon operating conditions (frequency and amplitude of the alternating magnetic field). \mnfe retains a greater magnetization change at high frequency, but its normalized $\chi''$ maximum is found at a lower angular frequency than the magnetite maximum even though its assumed $\tau_m$ is shorter. This demonstrates that the frequency-dependent response is not ordered by $\tau_m$ alone.  The frequency dependence is instead affected by the three relaxation parameters and by the energy-weighted electron, phonon, and magnon contributions that enter the susceptibility through Eq.~\eqref{eq:chi} and $G(\omega)$.

The amplitude of the magnetic field also affects the relaxation behavior. Around a frequency of 100 MHz, the magnetic damping in magnetite deviates most significantly from sinusoidal at large amplitudes. At 5 GHz, however, MnFe$_2$O$_4$ has the largest $|M_3|/|M_1|$ over the upper part of the sampled amplitude range so that the ordering by harmonic distortion changes with frequency. Cation redistribution produces smaller but measurable changes in $M$--$H$ trajectories, cycle work, entropy production, and subsystem temperatures, with the differences becoming most visible at larger amplitudes and higher frequencies. The cation configuration, thus, has a secondary influence relative to the family-level composition of the ferrite.

 The calculated $W_{\mathrm{cyc}}$ provides a mechanistic estimate of the energy density transferred through the modeled longitudinal quasiparticle-relaxation contribution. It is not a prediction of total core loss. The subsystem temperatures formulated here represent internal state variables rather than device steady-state temperatures. Within those limits, the formalism provides a physically resolved bridge between first-principles excitation spectra and engineering-scale magnetic models. The next step toward quantitative design use is clear: constrain the dynamic inputs experimentally, propagate their uncertainty through the coupled response, and then combine the intrinsic relaxation calculation with domain, electromagnetic, and heat-interactions descriptions of the actual core.

\FloatBarrier

\bibliography{references_rev5}
\end{document}